\documentclass[12pt]{article}
\usepackage{amsmath,amsthm,latexsym,amssymb,amsfonts,epsfig}

\makeatletter
\@addtoreset{equation}{section}
\makeatother

\def\be{\begin{equation}}
\def\ee{\end{equation}}
\def\ba{\begin{eqnarray}}
\def\ea{\end{eqnarray}}
\def\Nl{{\mathchoice
{\setbox0=\hbox{$\displaystyle\rm N$}\hbox{\hbox to0pt
{\kern0.4\wd0\vrule height0.9\ht0\hss}\box0}}
{\setbox0=\hbox{$\textstyle\rm N$}\hbox{\hbox to0pt
{\kern0.4\wd0\vrule height0.9\ht0\hss}\box0}}
{\setbox0=\hbox{$\scriptstyle\rm N$}\hbox{\hbox to0pt
{\kern0.4\wd0\vrule height0.9\ht0\hss}\box0}}
{\setbox0=\hbox{$\scriptscriptstyle\rm N$}\hbox{\hbox to0pt
{\kern0.4\wd0\vrule height0.9\ht0\hss}\box0}}}}
\def\Zl{{\mathchoice
{\setbox0=\hbox{$\displaystyle\rm Z$}\hbox{\hbox to0pt
{\kern0.4\wd0\vrule height0.9\ht0\hss}\box0}}
{\setbox0=\hbox{$\textstyle\rm Z$}\hbox{\hbox to0pt
{\kern0.4\wd0\vrule height0.9\ht0\hss}\box0}}
{\setbox0=\hbox{$\scriptstyle\rm Z$}\hbox{\hbox to0pt
{\kern0.4\wd0\vrule height0.9\ht0\hss}\box0}}
{\setbox0=\hbox{$\scriptscriptstyle\rm Z$}\hbox{\hbox to0pt
{\kern0.4\wd0\vrule height0.9\ht0\hss}\box0}}}}
\def\Ql{{\mathchoice
{\setbox0=\hbox{$\displaystyle\rm Q$}\hbox{\hbox to0pt
{\kern0.4\wd0\vrule height0.9\ht0\hss}\box0}}
{\setbox0=\hbox{$\textstyle\rm Q$}\hbox{\hbox to0pt
{\kern0.4\wd0\vrule height0.9\ht0\hss}\box0}}
{\setbox0=\hbox{$\scriptstyle\rm Q$}\hbox{\hbox to0pt
{\kern0.4\wd0\vrule height0.9\ht0\hss}\box0}}
{\setbox0=\hbox{$\scriptscriptstyle\rm Q$}\hbox{\hbox to0pt
{\kern0.4\wd0\vrule height0.9\ht0\hss}\box0}}}}
\def\Rl{{\mathchoice
{\setbox0=\hbox{$\displaystyle\rm R$}\hbox{\hbox to0pt
{\kern0.4\wd0\vrule height0.9\ht0\hss}\box0}}
{\setbox0=\hbox{$\textstyle\rm R$}\hbox{\hbox to0pt
{\kern0.4\wd0\vrule height0.9\ht0\hss}\box0}}
{\setbox0=\hbox{$\scriptstyle\rm R$}\hbox{\hbox to0pt
{\kern0.4\wd0\vrule height0.9\ht0\hss}\box0}}
{\setbox0=\hbox{$\scriptscriptstyle\rm R$}\hbox{\hbox to0pt
{\kern0.4\wd0\vrule height0.9\ht0\hss}\box0}}}}
\def\Cl{{\mathchoice
{\setbox0=\hbox{$\displaystyle\rm C$}\hbox{\hbox to0pt
{\kern0.4\wd0\vrule height0.9\ht0\hss}\box0}}
{\setbox0=\hbox{$\textstyle\rm C$}\hbox{\hbox to0pt
{\kern0.4\wd0\vrule height0.9\ht0\hss}\box0}}
{\setbox0=\hbox{$\scriptstyle\rm C$}\hbox{\hbox to0pt
{\kern0.4\wd0\vrule height0.9\ht0\hss}\box0}}
{\setbox0=\hbox{$\scriptscriptstyle\rm C$}\hbox{\hbox to0pt
{\kern0.4\wd0\vrule height0.9\ht0\hss}\box0}}}}
\def\Hl{{\mathchoice
{\setbox0=\hbox{$\displaystyle\rm H$}\hbox{\hbox to0pt
{\kern0.4\wd0\vrule height0.9\ht0\hss}\box0}}
{\setbox0=\hbox{$\textstyle\rm H$}\hbox{\hbox to0pt
{\kern0.4\wd0\vrule height0.9\ht0\hss}\box0}}
{\setbox0=\hbox{$\scriptstyle\rm H$}\hbox{\hbox to0pt
{\kern0.4\wd0\vrule height0.9\ht0\hss}\box0}}
{\setbox0=\hbox{$\scriptscriptstyle\rm H$}\hbox{\hbox to0pt
{\kern0.4\wd0\vrule height0.9\ht0\hss}\box0}}}}
\def\Ol{{\mathchoice
{\setbox0=\hbox{$\displaystyle\rm O$}\hbox{\hbox to0pt
{\kern0.4\wd0\vrule height0.9\ht0\hss}\box0}}
{\setbox0=\hbox{$\textstyle\rm O$}\hbox{\hbox to0pt
{\kern0.4\wd0\vrule height0.9\ht0\hss}\box0}}
{\setbox0=\hbox{$\scriptstyle\rm O$}\hbox{\hbox to0pt
{\kern0.4\wd0\vrule height0.9\ht0\hss}\box0}}
{\setbox0=\hbox{$\scriptscriptstyle\rm O$}\hbox{\hbox to0pt
{\kern0.4\wd0\vrule height0.9\ht0\hss}\box0}}}}

\title{{\sf Solving the Problem of Time}\\ {\sf in General Relativity 
and Cosmology} \\{\sf with Phantoms and k -- Essence}}
\author{{\sf T. 
Thiemann}\thanks{{\sf 
thiemann@aei.mpg.de,tthiemann@perimeterinstitute.ca}}\\
\\
{\sf MPI f. Gravitationsphysik, Albert-Einstein-Institut,} \\
           {\sf Am M\"uhlenberg 1, 14476 Potsdam, Germany}\\
\\
{\sf and}\\
\\
{\sf Perimeter Institute for Theoretical Physics,}\\ 
{\sf 31 Caroline Street N, Waterloo, ON N2L 2Y5, Canada}}
\date{{\sf\small Preprint AEI-2006-056}}

\begin{document}


\maketitle

\begin{abstract}
{\sf
We show that if the  
Lagrangean for a scalar field coupled to General Relativity only contains 
derivatives, then it is possible to completely deparametrise 
the theory. This means that\\
1.  
Physical observables, i.e. functions 
which Poisson commute with the spatial diffeomorphism and Hamiltonian 
constraints of General Relativity, can be easily constructed.\\
2. 
The physical time evolution of those observables is generated by  
a {\it natural} physical Hamiltonian which is (constrained to be) positive.

The mechanism by which this works is due to 
Brown and Kucha\v{r}. In order that the physical Hamiltonian is close to 
the Hamiltonian of the standard model and the one used in cosmology, 
the required Lagrangean must be that of a Dirac -- Born -- Infeld type. 
Such matter has been independently introduced previously by cosmologists
in the context of k -- essence due to Armendariz-Picon, Mukhanov and 
Steinhardt in 
order to solve the cosmological coincidence (dark energy) problem.
We arrive at it by 
totally unrelated physical considerations originating from quantum gravity. 

Our manifestly gauge invariant approach leads to important modifictaions 
of the interpretation and the the analytical appearance of the standard 
FRW equations of classical cosmology in the late universe. In 
particular, our concrete model implies that the universe should 
recollapse at late times on purely classical grounds.
}
\end{abstract}

\newpage

\section{Introduction}
\label{s1}

By ``the problem of time'' in General Relativity (GR) one means that 
GR is a completely parametrised system. That is, there is no natural 
notion of time due to the diffeomorphism invariance of the theory and 
therefore the canonical Hamiltonian which generates time 
reparametrisations vanishes. In fact, instead of a Hamiltonian there are 
an infinite number of spatial diffeomorphism and Hamiltonian constraints 
respectively, of which the canonical Hamiltonian is a linear combination, 
which generate infinitesimal spacetime 
diffeomorphisms\footnote{When the equations of motion hold.}. Physical
observables, sometimes called Dirac observables, are functions on phase 
space which are gauge invariant, that is, they Poisson commute with all 
constraints. In particular, they do not evolve with respect to the
canonical Hamiltonian. Hence ``nothing seems to happen in quantum 
gravity''.

The problem of time is not only of academic interest. One of the 
motivations for 
the present article actually comes from cosmology and can be phrased as 
the following question:\\
\\
{\bf Why is it that the FRW equations describe the physical time 
evolution which is actually observed for instance through red shift 
experiments, of physical, that is observable,
quantities such as the scale parameter?}\\
\\
The puzzle here is that these observed quantities are mathematically 
described by functions on the phase space which {\it do not 
Poisson commute with the constraints!} Hence they are not gauge 
invariant and therefore should not be observable in obvious 
contradiction to reality. Moreover, the time evolution 
described by the FRW equations is obtained from the Hamiltonian 
equations of motion generated by the Hamiltonian constraint and not 
by an actual Hamiltonian. This is due to the fact that the 
``Hamiltonian'' used to derive the FRW equations is actually 
constrained to vanish by one of the Einstein equations. The 
``evolution equations'' generated by a constraint must therefore be  
interpreted as gauge transformations and those, by the very definition 
of gauge transformations, are also not 
observable, again in sharp contradiction to observation. Thus we arrive at
the following devastating conclusion:\\
\\
{\bf Either the mathematical formalism, which has been tested 
experimetally so excellently in other gauge theories such as QED,
is inappropriate or we are missing some new physics.}\\
\\
We will show in this article that 
the problem of time and the above puzzle can be solved in the canonical 
approach to GR if one 
manages to deparametrise the theory. By this we mean that it is 
possible to write the 
Hamiltonian constraints in the form $C(x)=\pi(x)+H(x)$ where $\pi$ is the 
momentum conjugate to a scalar field $\phi$ and where $H(x)$ is a 
positive function
on phase space\footnote{As always in the canonical approach we 
assume that the spacetime manifold is diffeomorphic to $\Rl\times 
\sigma$ where $\sigma$ is an arbitrary three manifold and $x$ are local 
coordinates on $\sigma$.} which depends on neither $\phi$ or $\pi$. 
In this situation 
it is possible to construct explicitly physical observables and the 
function $H:=\int_\sigma \; d^3x\; H(x)$ is the {\it natural} physical 
Hamiltonian which generates the time evolution of those observables.
We will show explicitly that the scalar matter Lagrangean can be chosen 
in such a way that the physical Hamiltonian is close to the Hamiltonian 
of the standard model and the one used in cosmology and that 
the gauge invariant physical observables are closely related to the 
``non -- observables'' mentioned above. \footnote{Of course it is
conceivable that other than scalar matter can induce deparametrisation
while the corresponding Hamiltonian has the properties mentioned.
In particular, it would be desirable to find a scalar mode among the
gravitational degrees of freedom leading to deparametrisation. However,
this has proved to be impossible.}

The missing physics could  
therefore be a scalar matter component which in a precise sense is pure 
gauge. We therefore call it a {\it phantom field} because it is not 
directly observable. This is phenomenologically appealing because
scalar matter has not yet been observed in nature. Its main 
effect is that it provides a notion of physical time 
evolution, it is a {\it perfect physical clock}. Although it is pure 
gauge, its presence has further 
observable consequences: The physical Hamiltonian deviates slightly from 
the usual Hamiltonian that one uses in the standard model or cosmology 
and therefore changes the dynamics slightly. The associated modified 
dynamics of observable quantities can be used in principle in order to 
test a given, deparametrisation generating, model experimentally. 
In fact, the modified evolution equations generated by the physical
{\it Hamiltonian} rather than the Hamiltonian {\it Constraint} can be 
recasted into FRW form, however, at least for the concrete realisation 
of deparametrisation that we consider here, now the FRW 
equations
adopt additional terms which are {\it dynamically generated}. 
There are two types of modifications. The first type is expected: In 
the standard interpretation of the 
FRW equations, these can be interpreted as matter terms which at early 
times statisfy the equation of state of dust $w=0$ while at late times
it becomes a cosmological constant $w=-1$. However, the energy of 
the scalar field is negative which requires that there be positive 
energy 
matter with those equations of state in order to have overall positive
matter energy. Thus the model could be able to explain dynamically why there 
must be  dark matter and dark energy. The second type of 
modifications are deviations 
from the FRW form itself. At very late 
times, where ``late'' depends  
on the parameters of the model, the FRW interpretation breaks down and the 
universe evolves {\it drastically differently} with respect to the 
physical Hamiltonian. In fact, our concrete model suggests that the 
universe should recollapse on purely classical grounds.
Therefore, if we 
really observe evolution 
with respect to the physical time parameter corresponding to the 
physical Hamiltonian induced by our scalar field then the 
FRW equations are an approximation to the true evolution of the 
universe, which 
is valid at sufficiently early times of the universe only.
Of course, the parameters of the model and its dynamical constants of 
motion can be tuned such that the FRW equations are still valid today.
Let us finish this paragraph with the 
following provocative lesson:\\
\\
{\bf All textbooks on classical GR incorrectly describe the Friedmann
equations as physical evolution equations rather than what they really 
are, namely gauge transformation equations. 
The true evolution equations acquire possibly observable modifications 
to the gauge transformation equations whose magnitude depends on the 
physical 
clock that one uses to deparametrise the gauge transformation 
equations.}\\
\\
Both types of modifications just mentioned will of course not only 
happen in homogeneous cosmology but also in full GR. 
Notice that we do not exclude 
observable scalar matter such as an inflaton in the Lagrangean, rather 
we propose that 
whatever scalar or other {\it observable} matter is present in nature, 
there is in addition our negative energy scalar field which is actually 
the reason for 
why that other matter can mathematically be related to gauge invariant,
i.e. observable, quantities. In a sense, the mathematical formalism
(gauge theory) together with the experimental evidence (e.g. 
the experimental verification of the FRW equations) inavoidably force 
us to       
conclude that there is something like a negative energy matter field 
which therefore 
could be called a {\it prediction}\footnote{Of course, there may be 
other 
realisations of deparametrisation, different from a scalar field.
However, the conclusion that there is a matter component of which we are 
unaware when we treat the FRW equations as if they came from a true 
Hamiltonian rather than the Hamiltonian constraint, remains.}.

In this paper we show that it is possible to find a whole class 
of scalar field Lagrangeans with the required properties. The 
mechanism which leads to deparametrisation rests on a 
observation due to Brown and Kucha\v{r} made in their seminal 
work \cite{BK} which enabled 
them to reformulate the Hamiltonian constraints of GR such that they 
Poisson commute among each other, which is a necessary condition for  
deparametrisation as we will see. The only requirement is that 
the covariant scalar field Lagrangean depends only on 
the first derivatives of the scalar field.
However, it may nevertheless self -- interact 
due to a non -- polynomial Lagrangean similar to quintessence fields
\cite{Q} and more generally as in in k -- essence 
models\footnote{Basically, a k -- essence field is a scalar field 
$\Phi$ which 
depends non -- linearly on the kinetic term $g^{\mu\nu} \Phi_{,\mu} 
\Phi_{,\nu}$.}
due to Armendariz-Picon,
Mukhanov and Steinhardt \cite{K}. All possible 
mutually Poisson commuting Hamiltonian constraints 
have been found in \cite{M}, but only a subclass of them originate from 
a covariant Lagrangean which we will provide in this paper.
A, possibly unique, two parameter family within that class leads to 
physical Hamiltonians 
which approach the standard model Hamiltonian when the scalar field is 
close to being spatially homogeneous and that of standard cosmology at 
sufficiently early times. That it is spatially homogeneous 
(i.e. a 
constant) turns out to be a {\it natural} requirement in order that the 
scalar field defines a good (i.e. synchronised everywhere on 
$\sigma$) clock. 

Curiously, as we will see, this family of scalar field Lagrangeans,
to which we are driven naturally by physical and mathematical 
considerations, has been considered 
before by cosmologists \cite{Phantom} for entirely different reasons. Its 
physical properties 
agree
with what cosmologists call a phantom field\footnote{The generally 
accepted rough definition of a phantom field seems to be that in a 
cosmological setting the first order term in $g^{\mu\nu} \Phi_{,\mu} 
\Phi_{,\nu}$ of the Lagrangean comes with a coefficient which has a sign 
opposite to the sign in the  
Klein -- Gordon Lagrangean.}. It turns out that our family of  
Lagrangeans are necessarily of
Dirac -- Born -- Infeld type with a constant potential. 
Notice again that this phantom field is not directly observable. 
However, we can, and probably must in order to have a positive energy 
budget,
add further k -- essence matter.
Such observable k -- essence matter 
is being discussed as a candidate for dark energy and inflation by 
cosmologists.

We should mention that the deparametrisation 
technique is a special, very simple case of the more general 
``relational'' approach
due to Rovelli, see \cite{Rovelli} and references therein. The 
mathematical implementation of this idea has been much improved recently
\cite{Bianca} (see also \cite{TT}). It consists in choosing an infinite 
number of gauge fixing 
conditions called ``clocks'' and the afore mentioned physical 
observables 
are the gauge invariant extensions, off the associated gauge cut, of 
non -- invariant ``partial observables''. The 
analytical formulae are very complicated power series in general and there
are unsolved mathematical issues such as convergence of the series.
In contrast to the deparametrisation case, in the more general case the 
associated physical observables Poisson commute only weakly with the 
constraints, that is, when the constraints hold, they are weak Dirac 
observables. Observables coming from the deparametrised theory are strong 
Dirac observables which is mathematically much more convenient.
Fortunately, the much more complicated partial observable machinery is 
{\it not needed} in order to arrive at the results of the present paper.
All the results that we claim in this paper will be proved by elementary 
methods, the paper is self -- contained in that respect.

We emphasise that the formalism developed in this paper is {\it exact 
and 
non -- perturbative}. On the other hand, it is purely classical only so 
far. This is true for almost all the available literature on relational 
physics. In order 
to apply quantum theory to it, operator ordering issues 
have to be solved for the power series. This is a difficult issue 
in the general relational framework, however, under natural mathematical 
assumptions, we can actually solve the operator ordering problem as we 
will sketch in section \ref{s5}. Yet, it may be 
necessary to develop a perturbative scheme just like in S -- matrix 
theory.  This is a good approximation 
as long as the (kinematical) states with which we probe these 
observables are strongly 
peaked, at physical time $\tau$, at the phantom field value $\phi=\tau$
which explains why the phantom field should be close to spatially 
homogeneous. 
Hence, for a sufficiently short period period of physical time $\tau$, 
the approximation should be quite good.

With the formalism developed in this paper, a natural platform 
for carrying out 
cosmological quantum field theory\footnote{By this we mean Quantum 
Gravity in the sector whose classical limit is classical cosmology. This 
should not be confused with quantum cosmology which is just a quantum 
mechanical toy model of the actual quantum gravitational field theory.}  
within the framework of Loop Quantum Gravity (LQG) 
\cite{LQG} is launched. See \cite{BT} for a corresponding proposal.\\
\\
The article is organised as follows:\\
\\

As this article is intended for both cosmologists and quantum geometers,
in section two we state the results of our analysis without proofs. The 
proofs will be supplied in the remaining sections. Readers just 
interested in the results can therefore skip all the rest of the paper 
except for section seven. 
  
In section three we review the Brown -- Kucha\v{r} mechanism to generate 
mutually commuting Hamiltonian constraints.  

In section four we define the physical observables of the theory as 
well as the physical Hamiltonian originating from a general phantom field 
Lagrangean.

In section five we show that physical and mathematical considerations 
naturally lead to a Dirac -- Born -- Infeld scalar
field 
Lagrangean which for certain parameter range has the interpretation of 
what cosmologists call a phantom. The associated physical selection 
principle is that the corresponding
physical Hamiltonian is positive and close to that of the standard 
model (when the metric is flat).

In section six we derive the consequences of the gauge 
invariance principle for cosmology by computing and interpreting the 
modified FRW equations. 

In section seven we conclude and outline what we plan to do with our
formalism in the future, in particular in {\it quantum cosmology}.

\section{Summary}
\label{s0}

The scalar field Lagrangean which leads to deparametrisation, induces 
a positive Hamiltonian which is close to that used in the standard model 
when 
the metric is flat and 
which leads to physical equations of motion which are in agreement with 
the cosmological FRW equations 
is given by
\be \label{0.1}
L=-\beta+\alpha\sqrt{|\det(g)|}\sqrt{1+g^{\mu\nu}\Phi_{,\mu} 
\Phi_{,\nu}}
\ee
Here $\alpha,\beta$ are constants of dimension\footnote{We assume 
signature
$(-,+,+,+)$ and choose units for which $8\pi G_{{\rm Newton}}=1$.
Moreover, we assume that spatial coordinates are dimensionless while 
time coordinates have dimension cm and $ds^2=g_{\mu\nu} dX^\mu dX^\nu$
has dimension cm$^2$. We take $\Phi$ to have dimension cm so that 
the argument of the root in (\ref{0.1}) is dimensionfree.} cm$^{-2}$.
We must have necessarily $\alpha>0$ as we will see below.
The sign of $\beta$ is 
unconstrained. A natural value for $\beta$ would be $\beta=\alpha$ so 
that 
for small $(\nabla\Phi)^2$ the Lagrangean becomes $\alpha (\nabla 
\Phi)^2/2$ which up to the positive constant $\alpha$ is the massless 
Klein -- Gordon field Lagrangean with the wrong sign, i.e. it is a 
phantom. The other natural value for $\beta$ is 
$\beta=0$ because $\beta$ could always be absorbed into a cosmological 
term. Let us choose $\beta=0$ for concreteness in this preliminary 
discussion. Lagrangeans of the form (\ref{0.1}) are being discussed in k 
-- essence \cite{K}, albeit there with non -- trivial potential. For our 
purposes, non -- trivial potentials are forbidden. We arrive at the 
model (\ref{0.1}) by a totally independent mathematical and physical 
reasoning, namely deparametrisation, hence the fact that we stumble on k 
-- essence is rather curious. 

The canonical formulation leads to the following spatial diffeomorphism 
and Hamiltonian constraints respectively
\ba \label{0.2}
D_a^{{\rm tot}} &=& D_a+\pi \phi_{,a} 
\\
C^{{\rm tot}} &=& C-\sqrt{[1+q^{ab}\phi_{,a} 
\phi_{,b}][\pi^2+\alpha^2 \det(q)]} 
\nonumber
\ea  
Here $D_a,\;C$ respectively are the contributions to the spatial 
diffeomorphism constraint and the Hamiltonian constraint of the 
gravitational and non -- phantom matter degrees of freedom, $\pi$ 
is the momentum conjugate to $\phi$ and $q_{ab}$ is the metric intrinsic 
to the spatial slices with inverse $q^{ab}$. Clearly $a,b,..=1,2,3$ 
while $\mu,\nu,..=0,1,2,3$. From (\ref{0.2}) we see that $C$ is 
constrained to be positive. This will be important for what follows.
If we had chosen the other sign for $\alpha$ then the root in 
(\ref{0.2}) would come with the opposite sign and $C$ would 
be constrained 
to be negative. One can also not reverse the sign in front of 
$(\nabla\Phi)^2$ in (\ref{0.1}) because this would lead to a non -- 
definite argument of the root in (\ref{0.2}).

The interpretation of (\ref{0.2}) is that these are 
constraints, i.e. they must vanish. The canonical transformations on 
phase space that they generate are therefore not evolutions but gauge 
transformations. In fact, one can show \cite{Wald} that when the 
Einstein equations hold, the canonical transformations that they 
generate precisely coincide with spacetime diffeomorphisms. Any object 
which has non -- vanishing Poisson brackets with the constraints is 
therefore not observable because only gauge invariant objects have 
physical meaning. The problem of time is therefore that we do not have 
a priori a Hamiltonian which generates physical time evolution of gauge 
invariant objects.  

The Brown -- Kucha\v{r} mechanism \cite{BK} consists in the crucial 
observation 
that 
\be \label{0.3}
\pi^2 q^{ab}\phi_{,a} \phi_{,b}=q^{ab} D_a D_b=:D
\ee
when $D^{{\rm tot}}_a=0$. Thus, using (\ref{0.3}) we can solve 
$C^{{\rm tot}}=0$ for $\pi$ and obtain a different 
Hamiltonian constraint 
\be \label{0.4}
C^{\prime {\rm tot}}(x)=\pi(x)+
\left[ \sqrt{\frac{1}{2}[C^2-D-\alpha^2 Q]
+\sqrt{\frac{1}{4}[C^2-D-\alpha^2 Q]^2-\alpha^2 D Q}} 
\right](x)=:\pi(x)+H(x)
\ee
where $Q:=\det(q)$.
Together with the $D^{{\rm tot}}_a(x)$ it defines the same 
constraint 
surface\footnote{More precisely a subset of the full constraint surface.
There are altogether four components of the constraint 
surface corresponding to the four possible combinations of signs in 
front of the two square roots involved in (\ref{0.4}). While these 
components 
connect in lower dimensional submanifolds, each of them is preserved 
by the gauge transformations induced by the full Hamiltonian constraint.  
We therefore restrict from now on once and for all to the subset defined 
by $C^{\prime{\rm tot}}=0$.} as the 
system (\ref{0.2}) and is also first class. By virtue of
the Brown -- Kucha\v{r} mechanism, the new Hamiltonmian constraints 
even mutually Poisson commute among each other. The arguments of the 
roots in (\ref{0.4}) are constrained to be non -- negative as the 
derivation of that expression reveals.  

Since $H$ no longer depends $\pi,\phi$, we have managed to deparametrise 
General Relativity and in fact the quantity 
\be \label{0.5}
H:=\int_\sigma\; d^3x \; H(x)
\ee
is a positive Hamiltonian, it is not constrained to vanish and it is 
gauge invariant, it Poisson commutes with all constraints. Next let for 
any real number $\tau$ 
\be \label{0.6}
H_\tau:=\int_\sigma\; d^3x \; [\tau-\phi(x)]\;\;H(x)
\ee
Let $f$ be any spatially diffeomorphism invariant quantity on phase 
space which does not depend on $\phi$. Such functions are trivial to 
construct, a simple example is 
the volume $f=\int_\sigma\;d^3x \sqrt{\det(q)}$. Then the series
\be \label{0.7}
O_f(\tau):=f+\{H_\tau,f\}
+\frac{1}{2!} \{H_\tau,\{H_\tau,f\}\} 
+\frac{1}{3!} \{H_\tau,\{H_\tau,\{H_\tau,f\}\}\}+...
\ee
defines a one parameter family of gauge invariant function on phase 
space. Moreover, we have 
\be \label{0.8}
\frac{d O_f(\tau)}{d\tau}=\{H,O_f(\tau)\}
\ee
In other words, the map $\tau\mapsto O_f(\tau)$ describes the physical 
time evolution of gauge invariant objects generated by the Hamiltonian 
(\ref{0.5}).

The crucial additional property of $H$ in (\ref{0.4}) which was used 
in order to select the model (\ref{0.1}) is that when the scalar field
$\phi$ is spatially homogeneous, which is natural in order that it 
defines an everywhere (on $\sigma$) synchronised clock $\phi(x)=\tau$,
if the spatial diffeomorphism coinstraint holds and if $\alpha$ is 
sufficiently small then $H(x)\approx |C(x)|=C(x)$. Hence $H$ 
approximates the 
standard model Hamiltonian when the spacetime is close to being flat.
If we had chosen the other sign for $\alpha$ we would get 
$H(x)\approx |C(x)|=-C(x)$. Since $C(x)>0$ for usual matter when space 
is flat, it would follow that with this sign we cannot have flat space
and moreover that all matter contributions to the Hamiltonian come with 
the wrong (negative) sign. Hence the choice $\alpha>0$ is the only 
suitable one for our purposes. 

Let us investigate this more closely:\\
Since
$C=C^{{\rm grav}}+C^{{\rm s-matter}}+C^{{\rm ns-matter}}=0$ where
$C^{{\rm s-matter}}>0$ is the standard matter energy density, we must
have $C=C^{{\rm grav}}+C^{{\rm ns-matter}}<0$ where the latter
contribution is from non standard matter such as our scalar field. The
gravitational contribution $C^{{\rm grav}}$ is indefinite, there are
positive and negative scalar
modes contained in it and therefore it would be desirable to use a
negative
gravitational mode for deparametrisation. Unfortunately, such a mode
does not lead to deparametrisation because $C$ depends on
both the corresponding
$\pi,\phi$. Hence we consider $C^{{\rm ns-matter}}\not\equiv 0$. It
turns out that we must restrict on the portion of phase space where we
have $H\approx \pm |C^{{\rm grav}}+C^{{\rm s-matter}}|$, hence a priori
both signs in front of the square root in (\ref{0.4}) are allowed.
Hence we should have $C^{{\rm ns-matter}}<0$ or
$C^{{\rm ns-matter}}>0$ respectively in order that
$H\approx
C^{{\rm grav}}+C^{{\rm s-matter}}$ comes with the {\it positive} sign in
front of $C^{{\rm s-matter}}$ because on flat space this is the energy
density of standard matter. However, as we will see, if $C^{{\rm
ns-matter}}>0$ then the physical evolution equations adopt 
modifications which lead to a big rip
singularity in cosmological applications (the universe reaches infinite
size in a finite amount of time). Thus, if we want to avoid this, we are
naturally led to scalar matter with negative energy density.

If we would have $H(x)=C(x)$ exactly, then the physical evolution 
equations 
derived from $H$ for gauge invariant observables not involving 
$\phi,\pi$ would exactly
equal the gauge transformations on non -- gauge invariant quantities
not involving $\phi,\pi$ derived from the canonical ``Hamiltonian''
(with $N=1,\;N^a=0$)
\be \label{0.9}
H^{{\rm canon}}(N,\vec{N})=\int_\sigma\; d^3x \; [N(x)C(x)+N^a(x) 
D_a(x)]
\ee
which ignores the phantom field. This would justify 
why
the Hamiltonian constraint integrated against unit lapse is often used 
as a Hamiltonian. The phantom field, being pure gauge, would have 
absolutely no visible effect. However, since $H(x),\;C(x)$ do not 
exactly coincide, there are important modifications, both technically 
and conceptually, to which we turn now in a cosmological setting.\\
\\
Namely, we will see that the FRW equations must be provided with a new 
and gauge invariant interpretation. The actual physical evolution 
equations generated by the Hamiltonian leads to drastic 
modifications 
in the very late universe while in the early universe (including today) 
they keep their standard form to a very good approximation, 
depending on the numerical value of $\alpha$. In flat, homogeneous and 
isotropic models the FRW line element takes the form $ds^2=-dt^2+a(t)^2 
dx^a dx^b\delta_{ab}$ with scale factor $a$ and all constraints are 
identically satisfied due 
to the high symmetry of the Ansatz, except for a single Hamiltonian 
constraint
\be \label{0.10}
C^{{\rm tot}}=[-\frac{P^2}{12 a}+(\Lambda+\rho_m)a^3]+\rho_{{\rm 
phantom}} a^3=:C+\rho_{{\rm phantom}} a^3
\ee
where 
\be \label{0.11}
\rho_{{\rm phantom}}=-\sqrt{\frac{\pi^2}{a^6}+\alpha^2}
=:-\alpha\sqrt{1+x}
\ee
is the negative phantom energy, $P$ is the momentum conjugate to $a$,
$\Lambda$ is a cosmological constant and $\rho_m$ is the energy density 
of all non -- phantom matter. The important quantity
\be \label{0.12}
x:=\frac{\pi^2}{\alpha^2 a^6}
\ee
will be called the deviation parameter.
The phantom pressure is positive
\be \label{0.13}
p_{{\rm phantom}}=-\frac{1}{3a^2} 
\partial(a^3 \rho_{{\rm 
phantom}})/\partial a=\alpha\frac{1}{\sqrt{1+x}}
\ee
leading to an equation of state and speed of sound respectively
\be \label{0.14}
w_{{\rm phantom}}=
\frac{p_{{\rm phantom}}}{\rho_{{\rm phantom}}}=
-\frac{1}{1+x}=-
\frac{\partial p_{{\rm phantom}}/\partial x}{\partial \rho_{{\rm 
phantom}}/\partial x}=-c_{{\rm phantom}}^2
\ee
~\\
We can now do two, conceptually very different, things:\\ 
1.\\ 
First we
follow the standard procedure in cosmology. That is, we use the
constraint $C^{{\rm tot}}$ as if it was a Hamiltonian. The associated
equations of motions of non -- observable quantities such as the scale
factor then lead to the usual FRW equations. From the point of view of
gauge theory, the interpretation of those FRW equations as evolution
equations of observable quantities is, however, completely wrong. That
$a(t)$ is not observable, that is, not gauge invariant, can be easily
seen from the fact that $da(t)/dt=\{C^{{\rm tot}},a\}\not=0$. The 
correct
interpretation of those equations is that they describe the behaviour of
non -- observable quantities under the gauge transformations generated
by the Hamiltonian constraint. \\ 
2.\\ 
The second thing that we can do is
to compute the gauge invariant functions such as $O_a(\tau)$ using
(\ref{0.7}) with $H_\tau=(\tau-\phi)H$ where the Hamiltonian (\ref{0.6})
becomes
\be \label{0.15}
H:=\sqrt{C^2-\alpha^2 a^6}
\ee
and compute their physical evolution equations generated by 
(\ref{0.15}).\\
\\ 
Mathematically the two procedures are very similar to each other: 
In the first approach we compute $da/dt=\{C^{{\rm tot}},a\}$ and express 
$P$ in terms of $da/dt$. The first FRW equation then results 
by substituting $P$ in terms of $da/dt$ into the constraint equation 
$C^{{\rm tot}}$. Then we compute $dP/dt=\{C^{{\rm tot}},P\}$ and insert 
this into $d^2 a/dt^2=\{C^{{\rm tot}},\{C^{{\rm tot}},a\}\}$ which 
results in the second FRW equation. They take the usual form
\ba \label{0.16}
3(\frac{da/dt}{a})^2 &=& \Lambda+\rho_m+\rho_{{\rm phantom}} 
\\
3\frac{d^2a/dt^2}{a} &=& \Lambda-\frac{1}{2}[\rho_m+3p_m+
\rho_{{\rm phantom}}+3p_{{\rm phantom}}] 
\nonumber
\ea
In the second approach we compute 
$dO_a(\tau)=\{H,O_a(\tau)\}=O_{\{H,a\}}(\tau)$ and can then solve 
$O_P(\tau)$ in terms of $dO_a(\tau)$. The first FRW equation then 
results by 
expressing $C^{{\rm tot}}$ in terms of physical observables, that is,
computing $O_{C^{{\rm tot}}}(\tau)$ and imposing 
$O_{C^{{\rm tot}}}(\tau)=0$. That this should hold follows from
$d O_{C^{{\rm tot}}}(\tau)/d\tau=O_{\{H,C^{{\rm tot}}\}}(\tau)=0$ since
$H$ is an observable, hence $O_{C^{{\rm tot}}}(\tau)=O_{C^{{\rm 
tot}}}(\phi)=C^{{\rm tot}}=0$. The second FRW equation then is obtained
by computing $dO_P(\tau)/d\tau=O_{\{H,P\}}(\tau)$ and using this 
in $d^2 O_a(\tau)/d\tau^2=O_{\{H,\{H,a\}\}}(\tau)$. This results
in 
\ba \label{0.17}
3(\frac{dO_a/d\tau}{O_a})^2 &=& [\Lambda+O_{\rho_m}+O_{\rho_{{\rm 
phantom}}}](1+\frac{1}{x}) 
\\
3\frac{d^2 O_a/d\tau^2}{O_a} &=& 
\Lambda(1+\frac{4}{x})-\frac{1}{2}\{[O_{\rho_m}+O_{\rho_{{\rm phantom}}}]
(1-\frac{5}{x})+3 
[O_{p_m}+O_{p_{{\rm phantom}}}] (1+\frac{1}{x})\}
\nonumber
\ea
where now 
\be \label{0.18}
x=\frac{E^2}{\alpha^2 O_a(\tau)^6}
\ee
and where $E=H=-\pi$ is a constant of motion, namely the energy of the 
universe. \\
\\
Comparing (\ref{0.17}) and (\ref{0.18}) reveals:\\
1. \\
Although from the point of gauge theory it is incorrect to interpret 
the FRW equations (\ref{0.16}) as evolution equations of observable 
quantities, as long as $x$ is large, the actual physical evolution 
equations of observables (\ref{0.18}) generated by the physical 
Hamiltonian take exactly the same form. 
All that we have to do is to make the substitution $(t,a(t))\to 
(\tau,O_a(\tau))$.\\
2. When $x$ gets small, the correct equations (\ref{0.17}) differ 
drastically from the incorrect equations (\ref{0.16}). Notice that what 
we observe in experiment is really a gauge invariant object 
such as $O_a(\tau)$ and not $a(t)$. Of 
course, the concrete scenario for deparametrisation that we have 
proposed here may not be realised in nature, however, we insist that 
whatever matter is used for deparametrisation, there will be corrections 
to the standard FRW equations. This should have observable consequences!

Notice that we do not doubt the validity the Einstein equations 
(\ref{0.16}). They follow from the fundamental object $C^{{\rm tot}}$
which we also used in our construction. However, we stress that their 
interpretation as physical evolution equations of observables is 
fundamentally wrong. The domain of validity of the interpretation 
of the usual FRW equations as evolution equations is controlled 
by the deviation parameter $x$.
It depends on the kinematical model parameter
$\alpha$ and the dynamical constant of motion $E$. The 
critical value is $x=1$ and is reached at scale factor $O_a=\root 3 
\of{E/\alpha}$ which can be as large as we want for sufficiently small
$\alpha$. Thus we see that the mathematical formalism together with our 
concrete model predicts that the universe evolves differently at late 
times, that is, at large scale factor. We expect similar modifications
in other applications of GR such as black hole physics and it is an 
interesting speculation that the corresponding gauge invariant 
interpretation of Einstein's equations 
could predict large scale deviations from Newton's law which then could 
be in  
agreement with the measured rotation curves of galaxies. Notice that all 
of this is a purely classical effect, there is no quantum gravity 
involved in this although our motivation, deparametrisation, certainly 
comes from quantum gravity.
 
The fact that the phantom makes a negative contribution to the energy 
budget may be disappointing for supporters of k -- essence where 
$\rho_k>0$ is usually required. However, 
$\rho_{{\rm phantom}}<0$ is of  
no concern as long as the remaining matter makes an overall
positive contribution\footnote{The usual energy conditions on the 
energy momentum tensor do not make any restrictions on individual 
matter species but only on the overall matter content of nature.}. 
Actually, since the gravitational contribution to 
the 
Hamiltonian constraint is negative definite in cosmology, in fact
due to $C^{{\rm tot}}=0$ we must have $\rho_m>|\rho_{{\rm phantom}}|$.
Notice by the equation of state the phantom behaves like dust at 
small scales $\rho_{{\rm phantom}}\to -E/O_a^3$ and as a negative 
cosmological 
constant $\rho_{{\rm phantom}}\to -\alpha$ at large scales. This can be 
easily compensated by 
additional positive energy k -- essence matter or simply by ordinary
(dark?) matter plus an additional cosmological constant term 
$\Lambda-\alpha>0$. In a sense, if we want to explain the observational 
fact that the FRW equations describe the universe while their 
mathematical derivation violates the 
principles of gauge theory, then something like a phantom is needed for 
deparametrisation and in turn it requires something like k -- essence 
for reasons of 
total positive matter energy budget. From this point of view, both a 
phantom and k -- essence are a prediction of the mathematical formalism 
(gauge theory) together with observation (FRW cosmology). 

We will see 
furthermore that in order that the universe does not reach infinite size 
in finite $\tau$ time, it must in fact recollapse which can be achieved by 
a suitable choice of the parameters. Then the picture of a periodic 
universe 
arises if one can establish that Quantum Gravity effects avoid big bang 
and big crunch singularities. This would imply that the universe 
evolves through 
the ``would be'' singularities in an infinite number of cycles. Notice 
that 
recently \cite{A} a simple cosmological toy model has been rigorously 
quantised by the methods of LQG using precisely the gauge invariant 
programme suggested in \cite{BT} and for which possible classical 
foundations have 
been layed out in the present paper for the full theory. In that model,
the singularity is indeed quantum mechanically avoided which is a 
promising hint that in full 
LQG the singularity is avoided as well.\\
\\  
We close this section with some final remarks:\\
1.\\
From the point of view of a cosmologist nothing would be more natural 
than to use the scale factor itself as a clock: It is a monotonic 
function of the unphysical time parameter $t$ (until possible 
recollapse). Why did we not do that immediately (we can do it 
indirectly, see below)? There are two reasons. First of 
all, we wanted to provide a universal framework, i.e. to provide a 
physical notion of time in all possible situations and not only in 
homogeneous ones. However, in inhomogeneous situations, the notion of a 
scale factor is void. As a substitute one could consider the volume 
of (subsets of) $\sigma$ (the total volume is infinite for non compact 
topology of $\sigma$). However, this does not work for the same 
second reason for which also the scale factor itself is inappropriate
in the homogeneous situation: In order to achieve deparametrisation and 
to obtain a physical Hamiltonian with all the required properties, the 
clock variable(s) must be cyclic in the Hamiltonian constraint (at least 
weakly). This condition is violated for the scale factor and its 
inhomogeneous relatives due to the universal coupling of matter to 
gravity.\\
2.\\
Of course, {\it after one has deparametrised the system} and one is only 
dealing with physical quantities such as the physical scale factor 
$O_a(\tau)$ (or one of its inhomogeneous relatives) one can use it 
as a physical clock in place of $\tau$ itself
which is maybe better geared to what one does in reality.
One can then express time 
dependence of other observables $O_f(\tau)$ in terms of $O_a$ by solving 
$O_a(\tau)$ for $\tau$. 
In other words, while we cannot use the scale factor to deparametrise the 
system, we can still use it as physical clock {\it after} 
deparametrisation. The time evolution in terms of the physical scale 
factor will then be generated by a more complicated physical 
Hamiltonian.\\
3.\\
One could think that what cosmologists usually do in order to describe 
measurable quantities mathematically is actually precisely correct, that 
is ``relational''. For instance the redshift factor 
\be \label{0.19}
z(t_1,t_2):=\frac{\omega_1}{\omega_2}\approx \frac{a(t_2)}{a(t_1)}
\ee
is the ratio between the emission frequency $\omega_1$ of a 
spectral 
line (known from a table top experiment on Earth) and the absorption 
frequency $\omega_2$ observed on Earth is certainly measurable. Formula
(\ref{0.19}) relates this observable quantity to the ratio of the scale 
factors at unphysical emission time $t_1$ and absorption time $t_2$ 
respectively. We will now show that (\ref{0.19}) is in fact 
incorrect:\\ The reason is that the quantities $a(t)$ are not 
observable. In
order to see what is going on, we have to go through the derivation of 
the redshift formula. Consider a star at comoving distance $r$ from 
Earth. For light the geodesic is null and due to 
$ds^2=-dt^2+a(t)^2 dx^a dx^b\delta_{ab}$ we get as an equation of motion 
$a(t)\dot{r}(t)=1$. Formula (\ref{0.19}) then results from the fact that 
the beginning and the end of the wave travel the same comoving distance
$r=\int_{t_1}^{t_2} dt/a(t)=\int_{t_1+T_1}^{t_2+T_2} dt/a(t)$ with 
$\omega_j=2\pi/T_j$. This is certainly mathematically correct, however, 
the quantities $a(t_j)$ are not observable. In order to express $z$ in 
terms of observale quantities $O_a(\tau)$ we express the line element in 
terms of $\tau$ (see (\ref{4.45}))
\be \label{0.20}
ds^2=-d\tau^2 (1+\frac{1}{x}) +O_a(\tau)^2 dx^a dx^b 
\delta_{ab}
\ee
Notice that $\tau$ is no gauge parameter but a physical observable 
associated with the physical Hamiltonian, hence the factor $1+1/x$ cannot 
be transformed away by a diffeomorphism $\tau\mapsto \varphi(\tau)$ 
without changing the Hamiltonian. We now obtain the null geodesic equation 
of motion
$O_a(\tau) d O_r(\tau)/d\tau=\sqrt{1+1/x}$. The same argument now leads to 
the modified redshift factor relation
\be \label{0.21}
z(\tau_1,\tau_2)=\frac{\omega_1}{\omega_2}=
\frac{O_a(\tau_2)}{O_a(\tau_1)}\;
\sqrt{\frac{1+\frac{1}{x(\tau_1)}}{1+\frac{1}{x(\tau_2)}}},\;\; 
x(\tau)=\frac{E^2}{\alpha^2 O_a(\tau)^6}
\ee
and now all displayed quantities are observable. Hence we see that as 
long as $x$ is large, (\ref{0.21}) and (\ref{0.19}) agree in the 
following sense: What one incorrectly does in cosmology is to identify 
the unobservable gauge pair  
$(t,a(t))$ with the pbservable physical pair $(\tau,O_a(\tau))$. With this 
interpretation, 
the wrong relation (\ref{0.19}) is a good approximation to the correct 
relation (\ref{0.21}) as long as $x$ is large. However, there are 
large deviations especially in the late universe and of course the 
modification (\ref{0.21}) may have an observable effect on the 
interpretation of supernovae type Ia observations (standard candles) 
which provide evidence for recent accelarated expansion of the 
universe.\\
\\
\\ 
We now proceed to the mathematical and physical details.

\section{Review of the Brown -- Kucha\v{r} Mechanism}
\label{s2}

\subsection{Covariant, Minimally Coupled, Potential Free Scalar Fields} 
\label{s2.1}

In order to prepare for the explanation of the Brown -- Kucha\v{r} 
mechanism we review here the canonical formulation of general
scalar field Lagrangeans of a special class.\\
\\ 
We consider a general, covariant scalar field Lagrangean minimally 
coupled to the metric with action
\be \label{2.1}
S_{{\rm phantom}}=\int_M \; d^4X\; \sqrt{|\det(g)|} L(-g^{\mu\nu} 
\Phi_{,\mu} \Phi_{,\nu}/2)
\ee
where $L$ is an arbitrary function of the variable indicated. It will be 
crucial for the Brown -- Kucha\v{r} mechanism to work that the scalar 
field $\Phi$ only appears with derivatives, i.e. there is no non 
-- trivial potential 
term. Obviously, (\ref{2.1}) is 
invariant under Diff$(M)$.

As usual we perform a 3+1 split of the action \cite{Wald} and assume 
that $M$ is 
diffeomorphic to $\Rl\times \sigma$ where $\sigma$ is a three -- 
manifold of arbitrary topology. Hence, there is a foliation 
$t\mapsto \Sigma_t=Y_t(\sigma)$ of $M$ by spacelike hypersurfaces which 
are the images of $\sigma$ under a one parameter family of embeddings 
$t\mapsto 
Y_t$. This way we obtain a diffemorphism $\Rl \times \sigma\to M;\;
(t,x)\mapsto X:=Y_t(x)$. We consider the foliation vector field 
$T(X):=[\partial Y_t(x)/\partial t]_{Y_t(x)=X}$ which can be split as 
$T(X)=[N(t,x) n(X)+N^a(t,x) \partial Y_t(x)/\partial x^a]_{Y_t(x)=X}$.
Here $x^a,\;a=1,2,3$ are local coordinates of $\sigma$ while $X^\mu,\;
\mu=0,1,2,3$ are local coordinates of $M$. The vector field $n$ is 
everywhere normal to the foliation, that is, $g_{\mu\nu} n^\mu 
Y^\nu_{t,a}=g_{\mu\nu} n^\mu n^\nu+1=0$. The functions $N,N^a$ 
respectively are known as lapse and shift functions. 

We now pull back (\ref{2.1}) by the diffeomorphism $Y$ and express 
everything in terms of\\ $N(t,x),\; N^a(t,x),\; q_{ab}(t,x)=(Y_t^\ast 
g)_{ab}(x)$ and $\phi(t,x)=(Y_t^\ast \Phi)(x)$. It is not difficult to 
check that in the embedding coordinates the components 
of the metric tensor read $g_{tt}=-N^2+N^a N^b q_{ab},\;g_{ta}=q_{ab} 
N^b,\; g_{ab}=q_{ab}$ and for the inverse
$g^{tt}=-1/N^2,\; g^{ta}=N^a/N^2,\; g^{ab}=q^{ab}-N^a N^b/N^2$ where 
$q^{ac} q_{cb}=\delta^a_b$. It follows that (the lapse is assumed to be 
everywhere non -- negative)
\be \label{2.2}
\sqrt{|\det(g)|}=N\sqrt{\det(q)},\;\;
I:=-g^{\mu\nu} \Phi_{,\mu} \Phi_{,\nu}=(\nabla_n \phi)^2-q^{ab} 
\phi_{,a} \phi_{,b}
\ee
where $n=(T-N^a Y_{t,a})/N$ so that $\nabla_n \phi=(\dot{\phi}-N^a 
\phi_{,a})/N$.  

We are now in position to perform the Legendre transform. We have 
\be \label{2.3}
\pi(t,x):=\delta S_{{\rm phantom}}/\delta \dot{\phi}(t,x)=\sqrt{\det(q)} 
[\nabla_n \phi]
L'(I/2)
\ee
where the prime denotes the derivative with repspect to $I/2$.
From (\ref{2.3}) we infer
\be \label{2.4}
K:=[\frac{\pi}{\sqrt{\det(q)}}]^2=[L'(I/2)]^2 (I+V),\;\;
V:=q^{ab} \phi_{,a} \phi_{,b}
\ee
We assume that $L$ is such that (\ref{2.4}) can be solved uniquely for 
$I=J(K,V)$. Then (\ref{2.3}) can be solved for $\nabla_n \phi$
\be \label{2.5}
\nabla_n \phi= p/L'(J/2),\;p:=\pi/\sqrt{\det(q)}
\ee
We can now complete the Legendre transform
\be \label{2.6}
S_{{\rm phantom}}=\int_\Rl dt \int_\sigma d^3x (\pi \dot{\phi}-
[N^a \pi \phi_{,a} + N \sqrt{\det(q)} \{p^2/L'(J/2)-L(J/2)\}]
\ee
From (\ref{2.6}) we read off the contributions of the scalar field to 
the spatial diffeomorphism and Hamiltonian constraint respectively
\ba \label{2.7}
D_a^{{\rm phantom}} & = & \pi \phi_{,a}
\\
C^{{\rm phantom}} & = & \sqrt{\det(q)} [\frac{p^2}{L'(J/2)}-L(J/2)]
\nonumber
\ea

\subsection{The Brown -- Kucha\v{r} Mechanism}
\label{s2.2}

Let us denote by $D_a,\; C$ respectively the contribution 
to the spatial 
diffeomorphism and Hamiltonian constraint respectively of the 
gravitational field and all other matter fields (say of the standard 
model or one of its supersymmetric extensions). Then the spatial 
diffeomorphism constraint is given by $D_a^{{\rm tot}}=D_a+D_a^{{\rm 
phantom}}$ and 
the Hamiltonian constraint by $C^{{\rm tot}}=C+C^{{\rm phantom}}$.  
The simple, but crucial observation due to Brown and Kucha\v{r} is that
we may use the spatial diffeomorphism constraint in order to remove 
the dependence of $C^{{\rm tot}}$ on $\phi$ altogether, thus making it a 
function of $p$, the gravitational field and all the other matter 
fields only. Namely, we have, when $D^{{\rm tot}}_a=0$ 
\be \label{2.8}
V=q^{ab} \phi_{,a}\phi_{,b}=
\frac{q^{ab} D^{{\rm phantom}}_a  D^{{\rm phantom}}_b}{\pi^2}
=\frac{q^{ab} D_a  D_b}{\pi^2}
=\frac{1}{p^2} \frac{q^{ab} D_a  D_b}{\det(q)}
=: d/K
\ee
{\bf This is the Brown -- Kuchar Mechanism:} The field $\phi$, which 
appears only in the combination $V$ within the Hamiltonain constraint,
has been eleminated. This would not work 
if the Lagrangean also would depend on $\phi$ explicitly (potential term), 
not only through 
the combination $I$ which involves only derivatives of $\phi$.
  
Consider now the function 
$\tilde{J}(K,d):=J(K,V=d/K)$. Then, the 
Hamiltonian constraint can be equivalently described by the function
\be \label{2.9}
\tilde{C}^{{\rm tot}}=
\sqrt{\det(q)}[c+[\frac{K}{L'(\tilde{J}/2)}-L(\tilde{J}/2)]
\ee
where $C=\sqrt{\det(q)} c$.
Since the constraints form a first class system, also the new 
constraints do and they define the same constraint surface.

Notice that (\ref{2.9}) depends on $p$ only through $K$. We will now 
assume that we may solve (\ref{2.9}) for $K$ algebraically (possibly 
with several branches)
\be  \label{2.10}
K=G(c,d)
\ee
Notice that by construction $G$ is (constrained to be) non -- negative.
We may therefore write the Hamiltonian constraint in the still 
equivalent form 
\be \label{2.11}
C^{\prime {\rm tot}}=\pi+\sqrt{\det(q)} \sqrt{G(c,d)}
\ee
The other sign is also possible but the above choice leads to a 
positive physical Hamiltonian close to that of the standard model when 
the metric is flat (as mentioned in section (\ref{0.2}) we will 
restrict to the subset of the constraint surface defined by 
(\ref{2.11}) in what follows).

What is remarkable about the functions $[G(c,d)](x)$ is that they 
mutually Poisson commute among each other. The formal proof is 
as follows:
The constraints (\ref{2.11}) form a first class system, hence their 
mutual Poisson brackets is a linear combination of the constraints 
$C^{\prime {\rm tot}}=\pi+\sqrt{G(c,d)}, D^{{\rm tot}}_a$ with structure 
functions. However, since the $\pi$ Poisson commute among themselves as 
well as with $H=\sqrt{\det(q) G}$ because $G$ does not depend on $\phi$, 
it follows that $\{C^{\prime {\rm tot}}(x),C^{\prime {\rm tot}}(y)\}
=\{H(x),H(y)\}$ does not depend on $\pi,\phi$ any more. Thus, the linear 
combination of constraints with structure functions, which are non -- 
singular on the constraint surface, must conspire in 
such a way that the dependence of the bracket on $\pi,\phi$ drops out 
completely. Suppose then that 
\be \label{2.12}
\{C^{\prime {\rm tot}}(N),C^{\prime {\rm tot}}(N')\}=\int d^3x
[f_{N,N'}(x) C^{\prime {\rm tot}}(x)+  
f_{N,N'}^a(x) D^{{\rm tot}}_a(x)]  
\ee
where $N,N'$ are test functions and $C(N)=\int d^3x N(x) C(x)$.
Since (\ref{2.12}) does not depend on $\pi,\phi$ we may choose 
$\pi$ such that (\ref{2.11}) vanishes. Then only the second term in 
(\ref{2.12}) survives and must no longer depend on $\phi$. It follows 
that 
\be \label{2.13} 
\int d^3x  [f_{N,N'}^a(x) D^{{\rm tot}}_a(x)]_{\pi=-\sqrt{\det(q) G}}   
=\int d^3x [f_{N,N'}^a(x)]_{\pi=-\sqrt{\det(q) G}} (D_a-\sqrt{\det(q) G}
\phi_{,a})(x)  
\ee
We can now expand $g^a=(f_{N,N'}^a)_{\pi=-\sqrt{\det(q) G}}$ in 
powers of $\phi_{,a}$, that is
\be \label{2.14}
g^a=g^a_0+\sum_{n=1}^\infty g^{ab_1..b_n}_n \phi_{,b_1} .. \phi_{,b_n}
\ee
where the coefficients are supposed to be independent of $\pi,\phi$.
The resulting recursion relation is then given by 
($\pi=-\sqrt{\det(q)G}$ being understood)
\be \label{2.15}
D_a g^{ab_1..b_n}_n+\pi g^{b_1.. b_n}_{n-1}=0
\ee
and can be solved for instance by
\be \label{2.16}
g^{ab_1..b_n}_n=(-\frac{\pi}{h^c D_c})^{n-1} h^a h^{b_1} .. h^{b_{n-1}} 
g^{b_n}_0
\ee
where $h^a$ is an arbitrary function such that $h^a D_a\not=0$ and 
$g^a_0$ is also arbitrary. However, then 
\be \label{2.17}
g^a=g^a_0-\frac{h^a \pi g^b_0 \phi_{,_b}}{h^c[D_c+\pi\phi_{,c}]}
\ee
is singular on the constraint surface and in fact 
$g^a(D_a+\pi\phi_{,a})=g^a_0 D_a$ is not a linear combination of 
constraints.

We do not need to rely on such a formal argument: The rigorous proof 
is by actually computing the Poisson bracket. In \cite{M} we find the 
necessary and sufficient condition for expressions of the form 
$H=\sqrt{\det(q)G(c,d)}$ to be mutually Poisson commuting:
Consider functions of the form $H_n(Q,c,d):=Q^{n/2} h_n(c,d)$ where
$Q=\det(q)$. Then,
using the well known Poisson algebra generated by the Hamiltonian and 
spatial diffeomorphism constraints \cite{Wald} one can compute the 
Poisson brackets 
between the smeared functions $H_n(N):=\int d^3x N(x) H_n(x)$ and 
ask for the condition on $h_n$ such that $\{H_n(N),H_n(N')\}=0$.
This leads to the following first order partial differential equation
\be \label{2.18}    
\frac{n}{2} h_n \frac{\partial h_n}{\partial d} =
d [\frac{\partial h_n}{\partial d}]^2 
-\frac{1}{4} [\frac{\partial h_n}{\partial c}]^2
\ee
Dividing this equation by $h_n^2$ we get a PDE for $\ln(h_n)$ which one 
can solve by the method of separation of variables. The general 
solution, also called complete integral in the theory of first order 
PDE's, is given by the two parameter family
\be \label{2.19}
\ln[h_n(c,D;a,b)]=b+\frac{n}{4} \ln(d)+2\epsilon a c 
+\frac{\delta}{4}[2 s+n\ln(\frac{s-n}{s+n})]
\ee
where $s=\sqrt{n^2+16 a^2 d}$ and $\epsilon,\delta=\pm 1$. The so called 
general integral, which 
depends on an arbitrary function $g$, is obtained by solving the 
equation 
\be \label{2.20}
d \ln[h_n(c,d;a,b=g(a))]/da=
g'(a)+2\epsilon c +\delta \frac{s(a)}{2a}=0
\ee
for $a$ in terms of $c,d$ and to reinsert the solution $a=A(c,d)$ into 
$h_n$, that is, $h_n^g(c,d):=h_n(c,d;a=A(c,d),b=g(A(c,d)))$.\\
\\
Of course, in our case we are interested in the case $n=1$ and all we 
have to do is to check whether $h_1(c,d)=\sqrt{G(c,d)}$ solves 
(\ref{2.18}). The variety of solutions $h_1^g$ is certainly bigger than 
those that come from a covariant Lagrangean, that is, from a given 
function $L$. 

The reason for why we mention the variety $h_1^g$ is that 
we will be interested in solutions $h_1^g$ to (\ref{2.18}) with special 
properties and instead of guessing a suitable Lagrangean it may be
more constructive to select a candidate in the family $h_1^g$ and to 
ask from which Lagrangean, if any, it derives. In order to answer this 
question, suppose we have selected a solution $h_1^g(c,d)=\sqrt{G(c,d)}$ 
of
(\ref{2.18}). Then we solve (\ref{2.10}) $K=G(c,d)$ algebraically for 
$c=\tilde{c}(K,V)$
using again the key identity $d=V/K$. From this we infer 
$-\tilde{c}(K,V)=K/L'(J/2)-L(J/2)$ where $I=J(K,V)$ is the solution of 
$K=L'(I/2)^2 (I+V)$.
Hence we get  
\be \label{2.21}
\tilde{c}(L'(I/2)^2 (I+V),V)=L'(I/2)(I+V)-L(I/2)
\ee
In order that this be an ordinary first order differential equation for 
$L$ the explicit dependence of (\ref{2.21}) on $V$ must cancel. 
In order to achieve this, one performs algebraic manipulations to 
(\ref{2.21}) and obtains an equation which is a polynomial in $V$.
The coefficients of all powers of $V$ in that polynomial then have to 
vanish. This leads
in general to more than one ordinary differential equation for $L$
which are contradictory if $L$ does not exist and which are 
equivalent if $L$ exists. This 
is the necessary and sufficient condition for a given solution 
$\sqrt{G(c,d)}$ of (\ref{2.18}) to come from a covariant Lagrangean
of the form $L(I/2)$.

\section{Deparametrisation of General Relativity}
\label{s3}

The two ingredients that we need here are the following properties of 
the functions $H(x)=\sqrt{Q G(c,d)}(x)$ derived in the previous 
section:\\
1. they are mutually Poisson commuting.\\
2. they do not depend on $\pi,\phi$.\\
3. they are densities of weight $n=1$.\\
This is all we need in order to show that the following object
\be \label{3.1}
O_f(\tau):=[\alpha_M(f)]_{M=\tau-\phi},\;\;
\alpha_M(f)=\sum_{n=0}^\infty \frac{1}{n!} \{H(M),f\}_{(n)}
\ee
where $H(M)=\int_\sigma d^3x M(x) H(x)$ for some smearing function $M$,
is a one parameter family of strong Dirac observables, i.e. it Poisson 
commutes with both the 
spatial diffeomorphism constraint and the Hamiltonian constraint.
Due to the Poisson commutativity of $\phi(x),H(y)$, (\ref{3.1}) can also 
be written as 
\be \label{3.1a}
O_f(\tau)=\alpha_{\tau-\phi}(f)=\sum_{n=0}^\infty \frac{1}{n!} 
\{H_\tau,f\}_{(n)},\;\;
H_\tau:=\int_\sigma d^3x [\tau-\phi(x)] H(x)
\ee
The multiple Poisson bracket appearing in (\ref{3.1}), (\ref{3.1a})
is defined iteratively by $\{H_\tau,f\}_{(0)}=f$,
$\{H_\tau,f\}_{(n+1)}=\{H_\tau,\{H_\tau,f\}_{(n)}\}$. \\
The requirement on $f$ is that\\
A. it is already spatially diffeomorphism 
invariant\footnote{Such functions are easy to construct, any integral
of a scalar density of weight one constructed from the canonical 
variables is spatially diffeomorphism invariant.} and that\\
B. it does not depend on $\pi,\phi$.\\

For the expert the invariance of $O_f(\tau)$ under the gauge motions of GR
should be already obvious:\\ 
First of all, by inspection, $H_\tau$ is spatially diffeomorphism 
invariant because 
$[(\tau-\phi)H](x)$ is a scalar density of weight one. Hence 
$O_f(\tau)$ is spatially diffeomorphism invariant. Next, let us consider 
the Poisson automorphism
\footnote{For completeness, let us mention \cite{Bianca,TT} that if 
$f$ does depend on either $\pi$ or $\phi$ then (\ref{3.1}) must be 
generalised to $O_t(\tau):=[\beta_M(f)]_{M=\tau-\phi}$. For $f$ 
independent of $\pi,\phi$ we have $\beta_M(f)=\alpha_M(f)$, hence we 
arrive at (\ref{3.1}).}
$\beta_M(f):=\exp(X_{C^{{\rm tot}}(M)})\cdot f$
where $X_F$ denotes the Hamiltonian vector field of some phase 
space function $F$ and 
$C^{{\rm tot}}(M)=\int_\sigma d^3x M(x) C^{{\rm tot}}(x)$ for some test 
function $M$. Then we have 
\ba \label{3.1b}
\beta_M(O_f(\tau)) &=& \beta_M(\alpha_{\tau-\phi}(f))
\\
&=& \alpha_{\beta_M(\tau-\phi)}(\beta_M(f))
\nonumber\\
&=& \alpha_{\tau-\phi-M}(\beta_M(f))
\nonumber\\
&=& \alpha_{\tau-\phi-M}(\alpha_M(f))
\nonumber\\
&=& \alpha_{\tau-\phi}(f)=O_f(\tau)
\ea
where in the second step we used 
$\exp(X) \exp(Y)=\exp([\exp(X) Y \exp(-Y)]\exp(Y)$ for Hamiltonian vector 
fields $X,Y$, in the third we evaluated $\beta_M(H_\tau)=H_\tau-H(M)$, in 
the fourth we noticed that $\beta_M=\alpha_M$ on functions which do not 
depend on $\phi$ and in the last we noticed that $H_\tau, H(M)$ Poisson 
commute so that $\alpha_{\tau-\phi-M}=\alpha_{\tau-\phi}\circ 
\alpha_M^{-1}$. 

The next two subsections can therefore be skipped by the expert, for the 
non -- expert we will supply more details about symplectic geometry 
and will carry out the calculations in a less elegant but more pedestrian 
way which has the advantage of explicitly displaying the mechanism due 
to which all of this works.

\subsection{Invariance under Spatial Diffeomorphisms}
\label{s3.1}

In order to establish spatial diffeomorphism invariance, let us write
(\ref{3.1}) more explicitly as
\be \label{3.2}
O_f(\tau)=f+\sum_{n=1}^\infty \frac{1}{n!} \int_\sigma d^3x_1 ..
\int_\sigma d^3x_n \;(\tau-\phi(x_1))..(\tau-\phi(x_n)) 
\{H(x_1),..\{H(x_n),f\}..\}
\ee
where we have made use of the fact that $\phi(x)$ Poisson commutes with
$H(y),f$ by assumption. 

We will establish a result which is stronger than mere Poisson 
commutativity of 
$O_f(\tau)$ with $D^{{\rm tot}}(u)=\int_\sigma d^3x u^a(x) D^{{\rm 
tot}}_a(x)$ for arbitrary vector fields $u$ on $\sigma$: Let 
$s\mapsto \varphi^u_s$ be the one parameter family of 
spatial diffeomorphisms defined by the integral curves $c^u_x(s)$ which 
are the unique solutions of the system of ordinary differential 
equations $c^u_x(0)=x,\;\dot{c}^u_x(s)=u(c^u_x(s))$, that is: 
$\varphi^u_s(x):=c^u_x(s)$. We define for arbitrary functions $f$ on 
phase space 
\be \label{3.3}
\alpha_{\varphi^u_s}(f):=\sum_{n=0}^\infty \frac{s^n}{n!} 
\{D^{{\rm tot}}(u),f\}_{(n)}
\ee
Obviously $(\frac{d}{ds})_{s=0} \alpha_{\varphi^u_s}(f)=\{D^{{\rm 
tot}}(u),f\}$. It follows that $s\mapsto \alpha_{\varphi^u_s}$ is a one 
parameter family of Poisson automorphisms\footnote{That is, 
$\{\alpha(f),\alpha(g)\}=\alpha(\{f,g\})$ and 
$\alpha(f+g)=\alpha(f)+\alpha(g), \alpha(fg)=\alpha(f)\alpha(g),\;
\alpha(\bar{f})=\overline{\alpha(f)}$ for 
arbitrary, possibly complex valued, phase space 
functions $f,g$.} with the spatial diffeomorphism constraint as 
generator and one easily checks that 
$\alpha_{\varphi^u_s}(\tau-\phi(x))=\tau-\phi(\varphi^u_s(x))$ and  
$\alpha_{\varphi^u_s}(H(x))=|\det(\partial 
\varphi^u_s(x)/\partial x)| H(\varphi^u_s(x))$. 
Hence, the automorphisms generate transformations on phase space which
are in agreement with the fact that  
under a spatial diffeomorphism $\varphi\in {\rm Diff}(\sigma)$ the 
function $\tau-\phi(x)\mapsto \tau-\phi(\varphi(x))$ transforms as a 
scalar {\it because $\tau$ is a constant}. This would not hold if $\tau$ 
would be a non -- trivial phase space independent function. This will be 
important for our discussion below where we derive the relational 
origin of $O_f(\tau)$. On the other hand $H(x)\mapsto |\det(\partial 
\varphi/\partial x)| H(\varphi(x))$ transforms as a scalar density of 
weight one. 

Since $f$ Poisson 
commutes with $D^{{\rm tot}}(u)$ by assumption we have 
$\alpha_{\varphi^u_s}(f)=f$. Since 
$\alpha_{\varphi^u_s}$ is a Poisson automorphism it follows that 
\ba \label{3.4}
&&\alpha_{\varphi^u_s}([\tau-\phi(x_1)]..[\tau-\phi(x_n)] 
\{H(x_1),..\{H(x_n),f\}..\})
\\
&=& \alpha_{\varphi^u_s}(\tau-\phi(x_1)).. 
\alpha_{\varphi^u_s}(\tau-\phi(x_n)) 
\{\alpha_{\varphi^u_s}(H(x_1)),..\{\alpha_{\varphi^u_s}(H(x_n)),
\alpha_{\varphi^u_s}(f)\}..\})
\nonumber\\
&=& 
|\det(\partial \varphi^u_s(x_1)/\partial x_1)|..
|\det(\partial \varphi^u_s(x_n)/\partial x_n)|..
[\tau-\phi(\varphi^u_s(x_1))]..[\tau-\phi(\varphi^u_s(x_n))] 
\times
\nonumber\\
&& \times 
\{H(\varphi^u_s(x_1)),..\{H(\varphi^u_s(x_n)),f\}..\}
\nonumber
\ea    
Here we have used that the Jacobean $|\det(\partial 
\varphi/\partial x)|$ 
commutes with Poisson brackets. Now invariance of (\ref{3.2}) trivially 
follows since the Jacobean allows us to change variables under the 
integrals. 

This holds for the {\it infinitesimal} diffeomorphisms which 
are generated by $D^{{\rm tot}}_a$ and which allow us to explore the 
component of the identity of Diff$(\sigma)$. However, the invariance 
certainly extends to the full diffeomorphism group if $\phi,H$ transform 
as 
scalars of density weight zero and one respectively.

\subsection{Invariance under the Hamiltonian Constraint}
\label{s3.2}

Consider the smeared Hamiltonian constraint $C^{\prime {\rm tot}}(M)=
\int_\sigma d^3x M(x) C^{\prime {\rm tot}}(x)$ where $C^{\prime {\rm 
tot}}(x)=\pi(x)+H(x)$ and $M$ is an arbitrary test function. Using the 
explicit expression for $O_f(\tau)$ in 
the form (\ref{3.2}) we notice that there are contributions to 
$C^{\prime {\rm tot}}(M),O_f(\tau)\}$ from 
$\{C^{\prime{\rm tot}}(M),\phi(x)\}=M(x)$ and 
\be \label{3.5}
\{C^{\prime{\rm tot}}(M),\{H(x_1),..\{H(x_n),f\}..\}\}=
\int d^3x M(x) \{H(x),\{H(x_1),..\{H(x_n),f\}..\}\}
\ee
where we have used the fact that $f,H(x)$ and therefore 
$\{H(x_1),..\{H(x_n),f\}..\}$ do not depend on $\phi$.
We compute 
\ba \label{3.6} 
&& \{C^{\prime{\rm tot}}(M),O_f(\tau)\} =\{H(M),f\}+
\sum_{n=1}^\infty \frac{1}{n!} \int_\sigma d^3x_1 .. \int_\sigma d^3x_n 
\times \\
&& \times [- \sum_{k=1}^n [\tau-\phi(x_1)].. [\tau-\phi(x_{k-1})] M(x_k)
[\tau-\phi(x_{k+1}] .. [\tau-\phi(x_n)] \{H(x_1),..\{H(x_n),f\}..\}
\nonumber\\
&& +\int_\sigma d^3x_0 M(x_0) [\tau-\phi(x_1)]..[\tau-\phi(x_n)] 
\{H(x_0),..\{H(x_n),f\}..\}]
\nonumber
\ea 
The crucial Poisson commutativity of the $H(x_k)$ implies that the 
function\\ $(x_1,..,x_n)\mapsto \{H(x_1),..\{H(x_n),f\}..\}$ is completely 
symmetric in its arguments\footnote{Proof: We have 
$\{H(M_1),..\{H(M_n),f\}..\}=X_{H(M_1)}\cdot X_{H(M_2)} \cdot .. \cdot 
X_{H(M_n)} \cdot f$ where $H(M)=\int d^3x M(x) H(x)$ with arbitrary 
test functions $M_k$ and $X_{H(M)}$ is the Hamiltonian vector field of 
$H(M)$ which acts on functions by $X_{H(M)}\cdot f=\{H(M),f\}$. The 
relation $[X_{H(M)},X_{H(M')}]=X_{\{H(M),H(M')\}}$ holds between the Lie 
bracket of Hamitonian vector fields and the associated Poisson brackets. 
The claim now follows since $\{H(M),H(M')\}=0$.}. Thus, relabelling
$x'_0:=x_k,\;x'_1:=x_1,\;..,\;x'_{k-1}:=x_{k-1},\;x'_k:=x_{k+1},\;..,
x'_{n-1}:=x_n$ we can write (\ref{3.6}) in the form 
\ba \label{3.7} 
&& \{C^{\prime{\rm tot}}(M),O_f(\tau)\} = \{H(M),f\}
\nonumber\\
&&-
\sum_{n=1}^\infty \frac{1}{(n-1)!} \int_\sigma d^3x_0 .. \int_\sigma 
d^3x_{n-1} M(x_0)
[\tau-\phi(x_1)].. [\tau-\phi(x_{n-1})]
\{H(x_0),..\{H(x_{n-1}),f\}..\}
\nonumber\\
&& + \sum_{n=1}^\infty \frac{1}{n!}
\int_\sigma d^3x_0 .. \int_\sigma 
M(x_0) [\tau-\phi(x_1)]..[\tau-\phi(x_n)] 
\{H(x_0),..\{H(x_n),f\}..\}]
\nonumber\\
&=& 0
\ea 
which finishes the proof.

\subsection{Physical Hamiltonian}
\label{s3.3}

We claim that 
\be \label{3.8}
H:=\int_\sigma d^3x H(x)
\ee
is a physical Hamiltonian, that is, a physical observable which 
generates the time evolution $\tau \mapsto O_f(\tau)$ of all physical 
observables. 

That $H$ is spatially diffeomorphism invariant is trivial
because it is the integral of a scalar density of weight one. That it 
Poisson commutes with all the $C^{\prime{\rm tot}}(M)$ is also trivial 
because the 
$H(x)$ mutually Poisson commute among each other. Hence $H$ is a 
physical observable.

To see that $H$ generates the $\tau$ evolution we compute 
\ba \label{3.9}
&& \frac{d}{d\tau} O_f(\tau) 
= \sum_{n=1}^\infty \frac{1}{n!} \sum_{k=1}^n
\int_\sigma d^3x_1 .. \int_\sigma 
d^3x_n 
[\tau-\phi(x_1)]..[\tau-\phi(x_{k-1})]
[\tau-\phi(x_{k+1})]..[\tau-\phi(x_n)]
\times \nonumber\\
&& \times  \{H(x_1),..\{H(x_n),f\}..\}
\nonumber \\
&=& \sum_{n=1}^\infty \frac{1}{(n-1)!} 
\int_\sigma d^3x_0 .. \int_\sigma 
d^3x_{n-1} 
[\tau-\phi(x_1)]..[\tau-\phi(x_{n-1})] 
\{H(x_0),\{H(x-1),..\{H(x_n),f\}..\}
\nonumber\\
&=& \sum_{n=0}^\infty \frac{1}{n!} 
\{H,\int_\sigma d^3x_1 .. \int_\sigma 
d^3x_n 
[\tau-\phi(x_1)]..[\tau-\phi(x_n)] 
\{H(x-1),..\{H(x_n),f\}..\}\}
\nonumber\\
&=& \{H,O_f(\tau)\}
\ea
where we have made use of the same manipulations as in the previous 
section and 
in the last step we used again that $\{H(x),\phi(y)\}=0$. From 
(\ref{3.9}) it 
follows in particular that 
\be \label{3.10}
O_f(\tau)=\alpha_\tau(O_f),\;O_f:=O_f(0),\;\alpha_\tau(F):=
\sum_{n=0}^\infty \frac{\tau^n}{n!} \{H,F\}_{(n)}
\ee
which can be checked explicitly by expanding $O_f(\tau)$ in powers of 
$\tau$. In practice 
one therefore computes $O_f(\tau)$ by solving 
$df(\tau)/d\tau=\{H,f(\tau)\}$ and 
then chooses the ``constant'' of integration to be such that 
$f(\tau=\phi)=f$.

Finally, let us note the the identity
\be \label{3.12}
\{O_f(\tau),O_{f'}(\tau)\}=O_{\{f,f'\}}(\tau)
\ee
which immediately follows from the fact that $O_f(\tau)$ is the 
Hamiltonian flow of $f$ generated by $H_\tau$. Hence the physical 
observables satisfy a simple Poisson algebra if the $f,f'$ do.
Mathematically speaking, $f\mapsto O_f(\tau)$ is a one -- parameter 
family of Poisson homomorphisms.

\subsection{Relational Origin of the Formalism}
\label{s3.4}

The following section unveils the relational origin of our formalism and 
can be skipped by the reader merely interested in the physical 
application of the phantom field. On the other hand, one learns in this 
section why $\tau=const.$ rather than $\tau(x)$ is natural and why $H$
is a natural Hamiltonian. We also finish this section with some 
cautionary remarks which list some assumptions of the 
formalism which were not yet explicitly mentioned. \\
\\
Given a system of first class constraints $C_I$ possibly with structure 
functions, suppose we find functions $T_I$ such that the matrix $A$ with 
entries $A_{IJ}:=\{C_I,T_J\}$ is invertible and let $B=A^{-1}$. Consider 
the functions $C'_I:=\sum_J B_{IJ} C_J$, fix real values $\tau_I$ in the 
range of $T_I$ and let 
\be \label{3.11}
F^\tau_{f,T}=\sum_{\{n_I\}=0}^\infty 
\prod_I \frac{(\tau_I-T_I)^{n_I}}{n_I !} \prod_I X_I^{n_I} \cdot f
\ee
where $X_I$ denotes the Hamiltonian vector field of $C'_I$. One can show   
that the $X_I$ are weakly commuting, hence the sequence of the 
application of the $X_I$ in (\ref{3.11}) is irrelevant on the constraint 
surface \cite{Bianca}. One can show that (\ref{3.11}) is a weak Dirac 
observable, i.e. it Poisson commutes with all the $C_I$ on the 
constraint surface. One can also show \cite{TT} that the evolution in 
$\tau_I$
has a Hamiltonian generator $H_I(\{\tau_J\})$ which is defined via 
$\{H_I(\{\tau_J\}),F_{f,T}^\tau\}:=d F_{f,T}^\tau/d\tau_I$ for those 
$f$ which have vanishing Poisson brackets with the $T_I$ and their 
conjugate momenta.  
However, these Hamiltonians are not 
granted to be either positive or independent of the $\tau_J$ \cite{TT}. 
The physical 
meaning of $F_{f,T}^\tau$ is that it is 
the value of $f$ in the gauge\footnote{Notice that $T_I=\tau_I$ can be 
considered as a gauge fixing condition.} when $T_I$ assumes the value 
$\tau_I$. See \cite{Bianca} for more details.   

In General Relativity the label set of the $I$'s takes countably 
infinite cardinality and there are open issues with the convergence 
of (\ref{3.11}). In particular, the fact that (\ref{3.11}) is only a 
weak Dirac observable and $H_I(\{\tau_J\})$ a weak Hamiltonian is 
mathematically rather inconvenient. Moreover, the inversion of the 
matrix $A$ which is required in order to compute $F_{f,T}^\tau$ at least 
order by order is practically difficult for general choices of the 
$T_I$ which is 
why it is important to supply physical input towards choosing the 
``clocks'' $T_I$. Notice also that in General Relativity the $C_I$ will 
involve both the spatial diffeomorphism and the Hamiltonian constraint
even if $f$ is spatially diffeomorphism invariant because the 
Hamiltonian constraints do not close among themselves, they are 
proportional to a diffeomorphism constraint. This can be circumvented 
when the $T_I$ themselves are also spatially diffeomorphism 
invariant \cite{Bianca}, however, it is difficult to choose an 
algebraically 
independent set of such functions which also satisfy the requirement on 
$A$ and which can be considered as canonical configuration variables.

These comments reveal that (\ref{3.11}) is practically difficult to 
handle unless 
one manages to simplify it drastically, in particular the matrix $A$ 
should be simple. We claim that this is precisely what 
we managed to do in this paper: Let $b_I$ be an orthonormal basis of 
$L_2(\sigma,d^3x)$ such that $b_I$ also has finite $L_1(\Rl,d^3x)$ 
norm\footnote{For $\sigma=\Rl^3$ these could be Hermite functions.}.
Let $x\mapsto \tau(x)$ be an arbitrary function and define 
$\tau_I:=<b_I,\tau>,\;T_I:=<b_I,\phi>,\;\pi_I=<b_I,\pi>,
\;C'_I:=<b_I,C^{\prime{\rm tot}}>$
where we assume that $\tau,\phi,\pi,C^{\prime{\rm tot}}$ have at least 
finite sup norm on $\sigma$. Then we find for our $H_\tau$ 
that $H_\tau=\sum_I (\tau_I-T_I) 
(C_I-\pi_I)$. Notice that $A_{IJ}=\{C'_I,T_J\}=\delta_{IJ}$ is the unit 
matrix so that $C_I=C'_I$. Therefore, our $O_f(\tau)$ coincides with 
$F_{f,T}^\tau$ up to the fact that we use $\tau=const.$ rather than 
arbitrary $\tau(x)$. That $F_{f,T}^\tau$ Poisson commutes (even 
strongly) with the $C'_I$ is now a consequence of the fact that our $C'_I$ 
do close 
among themselves, namely they form an Abelean subalgebra of the 
constraint algebra. 

What is unclear however from the general relational framework  
is that this $F_{f,T}^\tau$ should Poisson 
commute with the spatial diffeomorphism constraint because  
formula (\ref{3.1}) does not involve the spatial diffeomorphism 
constraint at all
and the $T_I$ are not at all spatially diffeomorphism invariant.
Since 
$F_{f,T}^\tau$ coincides with $O_f(\tau)$ with $H_\tau$ replaced by 
$\int_\sigma d^3x (\tau(x)-\phi(x)) H(x)$ we see that this is 
automatically the case {\it if and only if} $\tau=const.$. This explains 
why, for gauge invariance reasons, $\tau=const.$ is the only reasonable 
choice for a function and 
why\footnote{This explains why we need $||b_I||_1<\infty$.} 
$\tau_I=\tau \int d^3x b_I$ is only a one parameter family of time 
evolutions rather than a ``bubble time evolution''. The generator of 
this $\tau$ evolution is then $\sum_I <1,b_I><b_I,H>=H(1)=H$ which
now has the advantage not to have an explicit $\tau$ dependence and 
which is positive by our construction.\\
\\ 
Thus we see that, by mere 
coincidence, the Brown -- Kucha\v{r} mechanism helps to {\bf drastically 
simplify 
the relational framework}. Rather than computing the 
infinite number of series (\ref{3.11}) and inverting complicated 
infinite dimensional matrices there is only one series (\ref{3.1}) and 
no matrix to invert. There is a distinguished notion of time generated 
by an invariant, positive and time independent physical Hamiltonian 
and the observables that we compute are strong observables, they have 
vanishing Poisson brackets with all constraints everywhere in the phase 
space, not only the constraint surface. Since there is only one 
series, convergence issues and, in quantum theory, operator ordering 
issues are much easier to settle. \\
\\
Remarks:
\begin{itemize}
\item[1.]
There is an issue which we have not considered so far: The functions 
$H(M),C^{\prime{\rm tot}}(M),\pi(M)$ may not converge for $M=\tau=const.$
if $\sigma$ is not bounded and/or 
may not be functionally differentiable if $\sigma$ has a boundary.
In fact, from the case of a massless Klein Gordon field which 
corresponds to the choice $L(I/2)=I/2$ and in case that the geometry is 
asymptotically flat one would assume that $\pi$ and therefore 
necessarily $H$
decay only as $1/r^2$ with respect to an asymptotic radial coordinate
and thus $H_\tau$ would blow up. 
We will assume that this problem is absent by a judicious choice of 
phantom model with corresponding fall -- off conditions for both $\pi$
and $H$. For instance we could use compact $\sigma$ without boundary and 
these issues would be absent. This is not sufficient for asymptically 
flat boundary conditions or in $k=0,1$ cosmology. In the model we derive 
in the next section this problem will be avoided automatically because 
the Hamiltonian turns out to have compact support, at least when the 
scalar field is close to homogeneous and the spatial diffeomorphism 
constraint holds. 
%
\item[2.]
One may worry that due to square roots which enter into the construction 
of $C^{\prime {\rm tot}}(M)$ its Hamiltonian vector field is singular or 
zero on 
the constraint surface. Indeed, that would be the case if one would 
drop the phantom field altogether and consider $H(x)$ as an alternative 
choice of the Hamiltonian constraint for GR and the remaining matter 
fields which together with the spatial diffeomorphism constraints $D_a(x)$
actually form a Lie algebra without structure functions \cite{M}. However, 
we do 
not drop the phantom field and due to the term $\pi(x)=C^{{\rm 
tot}}(x)-H(x)$ the Hamiltonian vector field does not vanish on the 
constraint surface and it is not singular because $H(x)$ is not 
constrained to vanish, rather $\pi(x)=-H(x)$.   
\item[3.]
The observables that we have constructed here are strong Dirac 
observables with 
respect to the new Hamiltonian constraints $C^{\prime{\rm tot}}(x)$ but 
since we have used the constraints in order to transform between this 
constraint and the original one, the observables will be only weak 
Dirac observables with respect to the original Hamiltonian constraints 
$C^{{\rm tot}}(x)$.
\item[4.]
We use the phantom field values $\phi(x)$ as physical clocks. Under the 
gauge transformations generated by the new Hamiltonian constraint 
$C^{\prime{\rm tot}}$ it transforms as $\delta\phi(x)=
\{C^{\prime{\rm tot}}(M),\phi(x)\}=M(x)$. Since the lapse function $M$ is 
required to be everywhere positive, it follows that under this unphysical
time evolution $\phi(x)$ evolves {\it strictly monotonously} for all $x$.
Therefore $\phi(x)$ is classically a perfect (i.e globally on phase 
space) clock for all $x$.
\end{itemize}

\section{Physical Scalar Field Models:\\ Dirac -- Born -- Infeld 
Phantom k -- Essence Lagrangeans}
\label{s4}

In the previous sections we have outlined why particular scalar field 
models in 
general allow to deparametrise GR. What is left to do is to exhibit a
(set of) 
model(s) which leads to a Hamiltonian that reduces to that of the standard 
model when the geometry is flat. This is the task of the present 
section. 
The analysis displayed here is far from 
complete and we will satisfy ourselves by finding one suitable model.
More general or improved models are left for future research.

\subsection{Selection of the Model} 
\label{s4.1}

The class of models we will look at is already constrained by the 
necessity to be able to solve (\ref{2.4}) 
\be \label{4.1}
K=[L'(I/2)]^2 \;(I+V)
\ee
algebraically for $I=J(K,V)$. Hence (\ref{4.1}) should lead to an 
algebraic equation at most of fourth order in $I$. We will restrict 
attention to models which only lead to quadratic equations in order to 
avoid the algebraic complications associated with Cardano's and Ferrari's
formulas for cubic and quartic equations resepctively. This restricts
us to functions $L$ that satisfy
\be \label{4.2}
[L'(I/2)]^2=\frac{a+b I}{\delta+\xi I+\zeta I^2}
\ee
where $a,b,\delta,\xi,\zeta$ are real constants such that the square roots 
that enter the integral $L$ of (\ref{4.2}), when evaluated at the 
solution $I=J$, of (\ref{4.1}) have positive arguments.

The integral of (\ref{4.2}) can be carried out for all values of 
$a,b,\delta,\xi,\zeta$ but in general involves 
complicated 
inverse trigonometric functions and logarithms. This would be bad 
because we also need to solve the Hamiltonian constraint for $K$ later 
on and such functions would lead to transcendental equations which 
cannot be solved algebraically. In order to avoid transcendental equations 
we have to specialise (\ref{4.2}) to one of the following cases:
\ba \label{4.3}
{\rm i.} && [L'(I/2)]^2=\frac{b^2}{4[a+bI/2]}
\\
{\rm ii.} && [L'(I/2)]^2=a^2
\nonumber\\
{\rm iii.} && [L'(I/2)]^2=(\frac{3}{2}b)^2 [a+bI/2]
\ea
which are readily integrated to
\ba \label{4.4}
{\rm i.} && L(I/2)=-(\beta+\epsilon \sqrt{a+bI/2})  
\\
{\rm ii.} && L(I/2)=-\beta + a I/2
\nonumber\\  
{\rm iii.} && L(I/2)=-(\beta+\epsilon \sqrt{a+bI/2}^3)  
\ea
where $a,b,\beta$ are real constants on which we will impose some 
restrictions in what follows and $\epsilon=\pm 1$. Remarkably, the 
Lagrangean i. in (\ref{4.4}) 
is precisely of the form of a Dirac -- Born -- Infeld type of phantom 
field with constant potential if we set $\epsilon=-1$, see e.g. 
the first reference in \cite{Phantom} and references therein. 

Let us first discuss case i. which turns out to be the right choice.
Equation (\ref{4.1}) now leads to 
\be \label{4.5}
K=\frac{b^2(I+V)}{4(a+bI/2)}\;\; \Rightarrow\;\;
J(K,V)=\frac{4aK-b^2 V}{b^2-2Kb}
\ee
It follows that 
\be \label{4.6}
a+bJ/2=\frac{a-bV/2}{1-2K/b}
\ee
which should be positive independently of the range of both $K,V$. Since
$K,V$ are manifestly positive, in order that the square root in 
(\ref{4.4}) is well defined (real valued) and in order that both sides of 
equation (\ref{4.3}) are positive we are forced to choose 
$a>0,\;b<0$. It is remarkable that this is possible although $J$ can take 
either sign, in fact $J$ is not bounded from below, however, it is 
bounded from above by $|2a/|b|$!

It follows that 
\ba \label{4.7}
\frac{C^{{\rm phantom}}}{\sqrt{\det(q)}} &=& \frac{K}{L'(J/2)}-L(J/2)
\\
&=& \epsilon \sqrt{a+Jb/2}[1-2K/b]+\beta
\nonumber\\
&=& \epsilon \sqrt{[a+Jb/2][1-2K/b]^2}+\beta
\nonumber\\
&=& \epsilon \sqrt{[a-bV/2][1-2K/b]}+\beta
\nonumber\\
&=& - \frac{C}{\sqrt{\det(q)}}=-c
\nonumber
\ea
where the last equality holds when the Hamiltonian constraint is 
satisfied. 

Using the Brown -- Kucha\v{r} key identity $V=d/K$
which holds when the spatial diffeomorphism constraint holds and taking 
the square of (\ref{4.7}) we find
\be \label{4.8}
(c+\beta)^2=a+d-\frac{bd}{2K}-\frac{2aK}{b}
\ee
This leads to a quadratic equation for $K$ which we can write as 
\be \label{4.8a}
(K-A)^2=A^2-B,\;\;A:=-\frac{b}{4a}[(c+\beta)^2-d-a],\;B:=d\frac{b^2}{4a}
\ee
Since the left hand side of (\ref{4.8a}) is positive, the right hand side 
of (\ref{4.8a}) is constrained to be positive as well. In order to 
to extend (\ref{4.8a}) off the constraint surface we write the two roots 
of (\ref{4.8a}) as\footnote{Lemma: For 
real numbers $x,y$ the equation $x^2=y$ implies 
$x^2=(y+|y|)/2$. \\
Proof: If $x^2=y$ holds then $y\ge 0$ so the second identity follows.
The second identity implies the first for $y\ge 0$, however it implies  
$x=0$ for $y<0$.}
\be \label{4.8b}
K=A\pm \sqrt{\frac{1}{2}(A^2-B+|A^2-B|)}
\ee
Since $K$ does not automatically vanish when $B=0$, only the positive sign 
in (\ref{4.8b}) is meaningful. Moreover, since $K$ is manifestly positive,
also the right hand side is constrained to be positive. We extend 
(\ref{4.8b}) off the constraint surface by 
\be \label{4.8c}
K=\frac{1}{2}\left[ A+\sqrt{\frac{1}{2}(A^2-B+|A^2-B|)}
+\left| A+\sqrt{\frac{1}{2}(A^2-B+|A^2-B|)} \right| \right]
\ee
which is now manifestly positive. If we want to avoid absolute values then 
we can write the solution explicitly in the form
\be \label{4.9}
K=G(c,d):=-\frac{b}{4a}[(c+\beta)^2-a-d]+
\sqrt{\{-\frac{b}{4a}[(c+\beta)^2-a-d]\}^2-\frac{d}{a}(b/2)^2}
\ee
This means in particular that we get the conditions
$A^2\ge B, (c+\beta)^2\ge a+d$ (recall that $a>0,\; b<0$). Hence 
also $(c+\beta)^2-(a+d)\ge 2 \sqrt{a} \sqrt{d}$, i.e.
\be \label{4.9a}
|c+\beta|\ge \sqrt{d}+\sqrt{a}
\ee
This condition, if imposed at all $x\in \sigma$, is gauge invariant and 
invariant under the physical evolution. To see this, notice that all 
quantities in (\ref{4.9a}) are spatial scalars, therefore (\ref{4.9a}) 
at $x$ is mapped to $\varphi(x)$ where it also holds by assumption. Next, 
all 
$H(x),\;x\in \sigma$ 
Poisson commute with $H(M)$ for all $M$ and with $H=H(M)_{M=1}$ in 
particular. Thus, if $H(x)^2=[\det(q)G(c,D)](x)$ is positive and 
meaningful for all $x\in 
\sigma$ then it will be in every gauge and under the physical time 
evolution. Thus, the extension off the constraint surface (\ref{4.8c})
which displays $H$ in manifestly positive form is only required fo the 
purpose of quantisation. For classical purposes, (\ref{4.9}) is 
completely sufficient since anyway we are only interested in the portion 
of phase space where (\ref{4.9a}) holds. 

Notice that for spatially homogeneous phantom fields $\phi$ the function 
$d$ is constrained to vanish. Hence, for $d=0$ (\ref{4.9}) reduces to
\be \label{4.10}
K=\frac{|b|}{4a}[(c+\beta)^2-a]+|(c+\beta)^2-a|] 
\ee
which is manifestly non negative and now there are no conditions 
on $c,d,\beta,a$.
We want this expression to coincide with $c^2$ for large $c$ in order 
that the Hamiltonian density equals $H(x)=C(x)=(\sqrt{\det(q)} c)(x)$ in 
this limit.
This enforces $b\approx -2a$ and $|\beta|,a$ small. Specifically we 
could 
set
$\beta^2=a$ in oder to remove the constant term in (\ref{4.10}) and 
$b=-2a$ in order that the coefficient of $c^2$ equals unity.
We may choose $a$ as small as we want in order to suppress the term 
linear in $c$ in (\ref{4.10}). For those values the Lagrangean 
becomes simply $L=-\sqrt{a}[\epsilon\sqrt{1-I}+ \delta]$ where $\delta=\pm 
1$. In case that $\sigma$ is not compact it makes sense to choose 
$\beta^2<a$ (e.g. $\beta=0$) because then (\ref{4.10}) vanishes in 
regions 
where $(c+\beta)^2\le a$. Hence, whatever the fall -- off conditions of 
the fields are at infinity, as long as they fall off at all, the 
support of (\ref{4.10}) will be compact, at least when 
the diffeomorphism constraint holds and when $\phi$ is homogeneous. 
Therefore, our model can be used, under the assumptions made, also if 
$\sigma$ is not compact and/or has a boundary. For other values of 
$\beta,a$
and in particular if there is a cosmological constant present, then the 
action converges anyway only if $\sigma$ is compact and we then also 
require $\partial\sigma=\emptyset$ in order to avoid boundary terms.

Once we have chosen $b=-2a, \beta=0$, (\ref{4.9}) reduces to
\be \label{4.9b}
K=\frac{1}{2}[c^2-a-d]\pm 
\sqrt{\{\frac{1}{2}[c^2-a-d]\}^2-ad}
\ee
It is then tempting to perform the limit $a\to 0$ 
so that (\ref{4.9b}) 
becomes
\be \label{4.9c}
K=\frac{1}{2}([c^2-d]\pm |c^2-d|)
\ee
which has the advantage to be manifestly positive even for $d\not=0$.
Since for $b=-2a=\beta=0$ the Lagrangean from which (\ref{4.9}) was 
derived
vanishes identically, (\ref{4.9c}) must come from a different 
Lagrangean. Indeed, it comes from the constrained, incompressible dust 
Lagrangean introduced by 
Brown and Kucha\v{r} in their original work \cite{BK}. 
Unfortunately, it is not of the form
$L(I/2)$, rather $L=-\frac{1}{2}\lambda(1-I)$ where $\lambda$ is a 
Lagrange multiplier field\footnote{In their work \cite{BK} they 
actually used $U_\mu=T_{,\mu}+W_a X^a_{,\mu}$ instead of 
$\Phi_{,\mu}$ 
where $W_a,X^a$ are additional six scalar fields and defined 
$I=-g^{\mu\nu} U_\mu U_\nu$. Then $\Phi:=T$ is a local clock field and 
$X^a,\;a=1,2,3$ are local position fields and together they are assumed 
to provide a local coordinate system of $M$. The functions $W_a$ are 
also Lagrange multipliers. We simply set $X^a,W_a=0$ here.}. The 
canonical 
formulation of this action leads to the contribution to the Hamiltonian 
constraint given by $C^{{\rm dust}}=\frac{1}{2}[\pi^2/(\lambda 
\sqrt{\det(q)})+(1+V)\sqrt{\det(q)}\lambda]$ and elimination of 
$\lambda$ 
by its equation of motion results in  
$C^{{\rm dust}}=\sqrt{\pi^2(1+V)}$. One can show explicitly by solving 
(\ref{4.9c}) for $c$ in order to obtain $C^{{\rm phantom}}$ that 
(\ref{4.9c}) cannot come from a Lagrangean of the form $L(I/2)$ by 
applying the procedure for the inverse problem mentioned at the end of 
section \ref{s2.2}. We therefore prefer scalar matter (\ref{4.9}) over the 
dust matter (\ref{4.9c}) which we feel to be awkward due to the Lagrange 
multiplier $\lambda$. What is awkward about this action is that the 
Euler -- Lagrange equations for $\lambda$ require $I=1$ and do not allow 
to solve for $\lambda$ while after Legendre transformation the 
constraint equation for $\lambda$ only can be solved by 
eliminating $\lambda$\footnote{If we compare this to the string, then 
this 
would be as if we would go from the covariant Polyakov string 
action to the Nabu -- Goto string action by 
eliminating the worldsheet metric (the analog of $\lambda$ here) by its 
equation of motion. However, both string actions are worlsheet 
covariant while this is not the case here.}. 
and to forget about $I=1$.\\ 
\\
Let us now consider the other cases. First of all, while
$a$ can be as small as we want in 
order to suppress unwanted terms in (\ref{4.9}), the 
case $a=0$ is singular. The case $a=0$, that is, 
$L(I/2)=\sqrt{I/2}$ or  $L(I/2)=\sqrt{I/2}^3$   
has to be treated independently. For $L(I/2)=\sqrt{I/2}$ we obtain   
an expression for the physical Hamiltonian which contains negative 
powers of $c,d$ which do not resemble the usual Hamiltonian. For 
$L(I/2)=\sqrt{I/2}^{3/2}$ as well as for case iii. in (\ref{4.3}) 
we are driven to an algebraic equation of ninth order for $K$ which can no 
longer be solved by quadratures, hence this model is ruled out for purely
mathematical reasons. Finally, 
the same analysis carried out for case ii. in (\ref{4.3})  
leads to $K=-c\pm\sqrt{c^2-d}$
which should be positive, hence $c$ is constrained to be negative.
Thus $C^{{\rm tot}}=\pi+\sqrt{\det(q)}\sqrt{-c+\sqrt{c^2-d}}$ and the 
Hamiltonian would be $H(x)=\sqrt{\det(q)}\sqrt{-c+\sqrt{c^2-d}}(x)$ 
which
for small $d$ becomes $\sqrt{2\det(q)|c|}$. Thus, the Klein Gordon 
Lagrangen produces the square root of $c$ which is also not what we 
want. \\
\\
Hence the only suitable model corresponds to case i. and we now proceed 
to explore its properties.

\subsection{Physical Properties of the Model}
\label{s4.2}

Let us first check that $h_1(c,d):=\sqrt{G(c,d)}$ of (\ref{4.9}) 
satisfies 
the PDE (\ref{2.18}). Since the calculation is not entirely trivial we 
display here some intermediate steps for the convenience of 
the reader. To simplify 
the computation we notice that  
$\sqrt{G}$ satisfies (\ref{2.18}) with $n=1$ if and only if $G$ satisfies 
(\ref{2.18}) with $n=2$. We find
\ba \label{4.11}
\frac{\partial G}{\partial d} &=& \frac{b}{4aR}(G-b/2)
\\
\frac{\partial G}{\partial c} &=& -\frac{b}{2aR} G (c+\beta)
\ea
where $R:=\sqrt{A^2-B},\;A:=b(a+d-(c+\beta)^2)/4a,\;B=(b/2)^2 d/a$.
Then (\ref{2.18}) with $n=2$ and $h_2:=G$ becomes  
\ba \label{4.12}
&& d [\frac{\partial G}{\partial d}]^2-\frac{1}{4} 
[\frac{\partial G}{\partial c}]^2-G[\frac{\partial G}{\partial d}]
\\
&=& (\frac{b}{4aR})^2[(d(G-b/2)-4aRG/b)(G-b/2)-G^2 (c+\beta)^2]
\nonumber\\
&=& (\frac{b}{4aR})^2[
\frac{4a}{b} G^2( \frac{b}{4a}[d-(c+\beta)^2]-R) +G(2aR-bd)+(b/2)^2 d]
\nonumber\\
&=& (\frac{b}{4aR})^2[
\frac{4a}{b} G(A+R)(A-R) -a G^2 +G(2aR-bd)+(b/2)^2 d]
\nonumber\\
&=& (\frac{b}{4aR})^2[
\frac{4a}{b} G B -a G^2 +G(2aR-bd)+(b/2)^2 d]
\nonumber\\
&=& (\frac{b}{4aR})^2[-a G^2 +2aRG +(b/2)^2 d]
\nonumber\\
&=& a(\frac{b}{4aR})^2[-G^2 +2RG +B]
\nonumber\\
&=& a(\frac{b}{4aR})^2[G(R+[R-G])+B]
\nonumber\\
&=& a(\frac{b}{4aR})^2[(R+A)(R-A)+B]
\nonumber\\
&=& 0
\ea
which is what we wanted to show.\\
\\
Next, in order to understand the meaning of $\epsilon$ and $\beta$ 
or $\delta$, we compute the equation of state 
of the model. The energy momentum tensor with our signature convention 
is given by
\be \label{4.10a}
T_{\mu\nu}=-\frac{2}{\sqrt{|\det(g)|}} \frac{\partial \sqrt{|\det(g)|} 
L(I/2)}{\partial g^{\mu\nu}}
=-[g_{\mu\nu}(\beta+\epsilon \sqrt{a+Ib/2})+\frac{\epsilon b}{2} 
\frac{\Phi_{,\mu} \Phi_{'\nu}}{\sqrt{a+b I/2}}]
\ee
Energy density and pressure become in the perfect fluid approximation 
$T_{\mu\nu}=\rho n_\mu n_\nu+p (g_{\mu\nu}+n_\mu n_\nu)$ (with respect 
to our 
unit timelike vector field $n$ normal to the foliation introduced in 
section \ref{s2}) 
\ba \label{4.10b}
\rho &=& T_{\mu\nu} n^\mu n^\nu=\beta+\epsilon\sqrt{(a-bV/2)(1- 2K/b)}
\\
p &=& \frac{1}{3}(\rho+g^{\mu\nu} T_{\mu\nu})=
-\frac{1}{3}(3(\beta+\epsilon)+\frac{\epsilon b}{2} 
\frac{V}{\sqrt{\frac{a-bV/2}{1-2K/b}}})
\ea
The equation of state ``field'' is defined by 
$w:=p/\rho$ and becomes for spatially homogeneous $\phi$ for which 
$V=0$
\be \label{4.10c}
w(y)=-\frac{\beta+\epsilon \sqrt{a}/y}{\beta+\epsilon 
\sqrt{a} y}
\ee
where $y=\sqrt{1-2K/b}\ge 1$. For $\beta=0$ we get $w=-1/y^2$ i.e 
$-1\le w \le 0$ independent of $\epsilon,a$. For $\beta\cdot \epsilon>0$ 
we also get $-1 \le w \le 0$. For $\beta\cdot 
\epsilon<0$
let $e=|\beta|/\sqrt{a}$. If $e>1$ then again $-1 \le w \le 0$. For
$e<1$, $w(y)$ has a maximum at $y_e=1/e+\sqrt{1/e^2-1}>1$ given by
$w(y_e)=(1/y_e)^2<1$ and $w(1)=-1,\; w(\infty)=0$ hence $-1\le w 
\le 1$. Finally for $e=1$ we get $0\le w(y)=1/y\le 1$.     

Next, the speed of sound, for spatially homogeneous $\phi$ is given by 
\be \label{4.10d}
c_s^2:=\frac{\partial p(\rho)}{\partial \rho}=
\frac{\partial p(y)/\partial y}{\partial \rho(y)/\partial 
y}=+\frac{1}{y^2}>0
\ee
independently of $\beta,a\epsilon$

Now if $\phi$ would be the only observable scalar field then we 
would need $\rho\ge 0,\;c_s^2>0$ for stability reasons, hence 
$\epsilon=+1$ in 
order that there are no restrictions on the range of $K,V$ and $\beta\ge 
-\sqrt{a}$ i.e. $e\le 1$. Then in order to get    
inflation and the recent accelarated expansion ($w<-1/3$) 
of the universe (dark energy) we must choose either $\beta\ge 
0$ or at least $-\sqrt{a}<\beta$. Since, however, the phantom field is 
pure gauge and has other purposes, it will  
not be associated with the physical inflaton and/or dark 
energy. Thus we keep the ranges of $\epsilon,a,b,\beta$ 
unrestricted up to the 
requirement that $a>0, b\approx -2a; \beta^2$ and $a$ small. We will 
see however, 
that the physical evolution equations of the next section still impose 
further restrictions.\\
\\
Let us set $b=-2a$ for definiteness. Then 
we conclude that we obtain the two parameter set of Dirac -- 
Born -- Infeld type Lagrangeans
\be \label{4.10e}
L=\sqrt{|\det(g)|}(-\beta+\alpha\sqrt{1+g^{\mu\nu}\Phi_{,\mu} 
\Phi_{,\nu}})
=\alpha\sqrt{|\det(g)|}(-1+\sqrt{1+g^{\mu\nu}\Phi_{,\mu} \Phi_{,\nu}})
+(\alpha-\beta)\sqrt{|\det(g)|}
\ee
where $\alpha:=-\epsilon\sqrt{a}\not=0$ is a real number with 
$|\alpha|$ small and 
$\beta$ is any real number with $|\beta|$ small. For small $I$
the first term in (\ref{4.10e}) becomes to linear order 
$\alpha\sqrt{|\det(g)|} g^{\mu\nu} \Phi_{,\mu} \Phi_{,\nu}/2$ which up 
to the constant $\alpha$ is just the usual Lagrangean for a 
massless 
Klein Gordon field. Thus, to this order the Lagrangean has the correct 
sign in 
front of the kinetic term for $\alpha<0$ and it becomes a phantom in the 
cosmological sense for $\alpha>0$. The second term represents a 
contribution by $\alpha-\beta$ to the cosmological constant which is a 
positive 
contribution for $\alpha-\beta$ negative. Since the cosmological term 
can always 
be absorbed into the contribution $C$ to the Hamiltonian constraint by 
gravity and the remaining (physical) matter, a natural choice would be
$\beta=\alpha$ which would then imply $e=1$ hence $0\le w\le 1$. \\
\\
Summarising, we have found a simple scalar field model which in 
the physical situation of interest, that is, a roughly homogeneous 
phantom field in order that $\phi(x)\approx 
\tau=const.$ is a good approximation for a physical clock, 
gives rise to a satisfactory physical Hamiltonian. It decays 
sufficiently fast at spatial infinity (in fact has compact support) 
when simultaneously the spatial diffeomorphism constraint holds. 
Decay requirements can be avoided if there is a non -- 
vanishing cosmological constant so that $\sigma$ is compact and we 
impose also $\partial\sigma=\emptyset$ in this case to avoid boundary 
terms.

It would be interesting 
to improve the model in order that it is applicable also in situations 
where $\sigma$ is bounded and has a boundary. Also, for mathematical 
reasons in 
particular in view of quantisation (see the next section) it would be more 
convenient to
have a manifestly positive Hamiltonian. A possible starting point is the 
manifestly positive extension (\ref{4.8c})
off the positivity constraint surface. Finally it might be possible 
to find a model such that $H$ approximates $C(N=1)$ even when $V$ is not 
close to zero. One way to investigate this would be to solve the 
inverse problem
mentioned at the end or section \ref{s2.2}.
We leave this to future research.

\section{Consequences for Cosmology}
\label{s4.3}

The Hamiltonian constraint for our model is given by\\
$C^{\prime{\rm tot}}(x)=\pi(x)+\sqrt{\det(q)G(c,d)}(x)=:\pi(x)+H(x)$ and 
we see that the Hamiltonian becomes for small $V$ and large $c(x)^2\gg 
\alpha^2,\beta^2$ 
approximately $H=\int_\sigma d^3x C(x)$ which is just the integrated
contribution, of the gravitational and non -- phantom like matter degrees 
of freedom, to the original Hamiltonian 
constraint.
It would result from the canonical Hamiltonian   
by choosing the lapse to equal unity, the shift to equal zero and 
by dropping the phantom field 
contribution from $C^{{\rm tot}},\;D_a^{{\rm tot}}$. This explains why 
in 
the presence of our particular phantom 
field model chosen, evolution with unit lapse and zero shift 
with respect to the canonical, original ``Hamiltonian'' $H^{{\rm 
canon}}(N,\vec{N})=C^{{\rm tot}}(N)+D^{{\rm tot}}(\vec{N})$ 
of functions on phase space not involving $\phi$ {\it approximately} 
equals  
the physical evolution of the non -- phantom like degrees of freedom.\\
\\
We will now illustrate the meaning of ``approximately'' in the context 
of isotropic and homogeneous minisuperspace models, that is, FRW 
cosmology. The presence of the phantom field, while not directly 
observable, will still have an important impact on the conceptual 
(interpretation) and technical (matter content) aspects of the FRW 
equations as well as on their validity. Namely, we will see that the 
FRW equations are only an approximation to the actual physical time 
evolution of observable quantities generated by the physical 
Hamiltonian. This also serves to explain the 
formalism in a simple context without the field theoretic 
complications\footnote{Notice that as usual we quotient all equations 
by the infinite coordinate volume $\int d^3x$ in the $k=0,-1$ models.}.\\
\\
We consider the experimentally almost confirmed case of a spatially 
flat ($k=0$) model for which the usual FRW line element is given by
$ds^2=-dt^2+a(t)^2 \delta_{ab} dx^a dx^b$ where $a$ is called the scale 
factor. Comparing with the general 
ADM line element $ds^2=-[N^2+q_{ab} N^a N^b]dt^2+2q_{ab} N^b 
dx^a+q_{ab} dx^a dx^b$ we read off $N=1,\;N^a=0,\;q_{ab}=a(t)^2 
\delta_{ab}=:Q \delta_{ab}$. Here we work with dimensionless 
spatial coordinates $x^a$ so that $a$ has dimension cm while the 
unphysical time (or foliation parameter) $t$ has dimension cm. We also
take our scalar field to have dimension cm so that $I=(d\phi/dt)^2$ is 
dimensionless. 

We begin by specialising the canonical formulation of GR to isotropic 
and homogeneous situations. The extrinsic curvature 
$K_{ab}=(\dot{q}_{ab}-{\cal 
L}_{\vec{N}} q_{ab})/(2N)$ where $\cal L$ denotes the Lie 
derivative reduces to $K_{ab}=a\; da/dt \delta_{ab}$, hence the 
momentum $P^{ab}=\frac{1}{2}\sqrt{\det(q)}[q^{ac} q^{bd}-q^{ab} 
q^{cd}]K_{cd}$ conjugate\footnote{Recall that the gravitational and 
cosmological action 
are multiplied by $1/(16\pi G)=1/2$ which explains the factor $1/2$ in 
front here.} to $q_{ab}$ becomes 
$P^{ab}=-\dot{a} \delta^{ab}=:P_Q \delta^{ab}/3$. Here $Q,P_Q$ are 
canonically conjugate. The canonical transformation from 
$(Q=a^2,P_Q)$ to $(a,P=2a P_Q)$ reveals $P=-6 a\;da/dt$.
The spatial diffeomorphism constraint vanishes 
identically if spatial homogeneity is assumed and the 
contribution to the Hamiltonian 
constraint of the gravitational degrees of freedom and the cosmological 
term becomes 
\be \label{4.13}
C^{{\rm grav}}=\frac{1}{2}(\sqrt{\det(q)}[(q^{ac} 
q^{bd}-q^{ab} 
q^{cd}) K_{ab} K_{cd}-
R^{(3)}]+2\Lambda\sqrt{\det(q)})
=-\frac{P^2}{12 a}+\Lambda a^3
\ee
where $R^{(3)}$ is the curvature scalar of $q_{ab}$ which vanishes 
identically.   

For ordinary matter we will make as usual a perfect 
fluid Ansatz for the energy 
momentum tensor $T_{\mu\nu}=\rho_m n_\mu n_\nu+p_m (g_{\mu\nu}+n_\mu 
n_\nu)$ with $n^\mu=\delta^t_\mu$
where energy density $\rho_m$ and pressure are related by 
$-3 a^2 p_m=d(\rho_m a^3)/da$. For instance for a Klein - Gordon field
$\rho_m=(\pi_m^2/a^6+U)/2$ where $U$ is its potential and thus
$p_m=(\pi_m^2/a^6-U)$.  
Since in general $c^{{\rm grav}}=-2[G_{\mu\nu}+\frac{1}{2}\Lambda 
g_{\mu\nu}]n^\mu n^\nu$ we find that the contribution to the Hamiltonian 
constraint of gravity and non -- phantom matter reads in 
general $C=C^{{\rm grav}}+a^3 \rho_m$.  
If there is a phantom present, as we 
advertised in the present paper then the 
total Hamiltonian constraint reads $C^{{\rm tot}}=C+a^3 
\rho^{{\rm phantom}}=C+C^{{\rm phantom}}$.

As we showed in section (\ref{s4.1}) the phantom field contribution to 
the Hamiltonian constraint is given by (remember $b=-2a,\;V=0$)
\be \label{4.14}
\rho^{{\rm phantom}}=C^{{\rm phantom}}/a^3=
\beta-\epsilon\sqrt{K+\alpha^2} 
=\beta-\epsilon\sqrt{\pi_{{\rm phantom}}^2/a^6+\alpha^2}
=:\beta-\epsilon\alpha\sqrt{1+x}
\ee
where we now take $\alpha>0,\;\epsilon=\pm 1$ and have introduced the 
``deviation parameter''
\be \label{4.15}
x:=\frac{E^2}{\alpha^2 a^6}
\ee
which will be crucial for what follows. We noticed that the gauge 
invariant quantity  
$\pi_{{\rm phantom}}$ is also a constant of the physical motion 
because the physical Hamiltonian $H=-\pi_{{\rm phantom}}$ does not 
involve $\phi$ (it is cyclic) and 
therefore we have denoted the energy squared constant of 
motion by $E^2:=\pi_{{\rm phantom}}^2$.  
Since (\ref{4.14}) has the same dimension as the cosmological constant 
$\Lambda$ we see that $\alpha,\beta$ have 
dimension 
cm$^{-2}$, $\pi,E$ have dimension cm and $x$ is dimensionless. The 
phantom pressure is given by
\be \label{4.15a}
p_{{\rm phantom}}=-\frac{1}{3a^2}d(a^3 \rho_{{\rm phantom}})/da
=-\beta+\epsilon \alpha^2 a^3/\sqrt{E^2+\alpha^2 a^6} 
=-\beta+\epsilon \frac{\alpha}{\sqrt{1+x}}
\ee
This gives an equation of state 
\be \label{4.16}
w_{{\rm phantom}}=\frac{p_{{\rm phantom}}}{\rho_{{\rm phantom}}}=
-\frac{\beta-\epsilon\alpha/y}{\beta+\epsilon\alpha y}
\ee
with $y=\sqrt{1+x}$. The phantom speed of sound is given by
\be \label{4.17}
c^2_{{\rm phantom}}=\frac{dp_{{\rm phantom}}/dy}{d\rho_{{\rm 
phantom}}/dy}=\frac{1}{y^2}
\ee
which is always positive independently of $\beta,\alpha,\epsilon$. 

The canonical ``Hamiltonian'' of the theory without the phantom is 
$H^{{\rm canon}}=C$. Without the phantom, however, it is constrained to 
vanish and 
therefore should not be interpreted as a Hamiltonian, rather as a 
Hamiltonian constraint which generates gauge transformations and not 
any observable evolution. With the phantom, $C$ is not constrained to 
vanish. The {\it physical Hamiltonian}, with the phantom 
present then follows from (\ref{4.10}) with $b=-2a$  (remember $d=0$
identically in exactly homogeneous cosmology)
\be \label{4.18}
H=a^3\sqrt{\frac{1}{2}[(c+\beta)^2-\alpha^2]+|(c+\beta)^2-\alpha^2|]} 
\ee
which approaches $|C|=a^3 |c|$ for $|c| \gg |\beta|,\alpha$. It 
vanishes 
identically when $|c+\beta|\le \alpha$ (this never happens on the 
constraint surface). Hence, as long as $c>0$ and 
$c\gg 
|\beta|,\alpha$ the physical Hamiltonian $H$, in presence of the 
phantom, is 
in good agreement with the canonical Hamiltonian constraint, in absence of 
the phantom. 

We will now derive the FRW equations from the canonical formalism and 
compare them with the evolution equations of physical observables. We 
begin with the
standard FRW equations which consist of a set of two equations. The 
first one 
is just the constraint equation $C^{{\rm tot}}=0$ with $P$ eliminated 
by the 
equation of motion for $a$ which gives a condition on $(da/dt)^2$. The 
second one involves $d^2a/dt^2$ and is obtained by eliminating $dp/dt$ 
by its equation of motion. Notice that by ``equation of motion'' we mean 
actually gauge transformations generated by $C^{{\rm tot}}$. On the 
other hand, we 
can compute evolution equations generated by the physical Hamiltonian 
$H$.
\begin{itemize}
\item[1.] {\it Gauge Transformations generated by $C^{{\rm tot}}$: 
Standard FRW Equations}\\
The gauge transformation for $a$ is
\be \label{4.19}
\frac{da}{dt}:=\{C^{{\rm tot}},a\}=\frac{\partial C}{\partial 
p}=-\frac{P}{6a}
\ee
Inserting (\ref{4.19}) into the constraint equation $C^{{\rm tot}}=0$ we 
therefore 
find the first one of the FRW equations
\be \label{4.20}
3 (\frac{da}{dt})^2/a^2=\Lambda+\rho_m+\rho_{{\rm phantom}}
\ee
The second equation is obtained by solving (\ref{4.20}) for $da/dt$,
taking the second derivative and using the conservation law
$d\rho/dt+3(\rho+p) da/dt=0$ which gives
\be \label{4.24}
3\frac{d^2 a}{dt^2}/a
=\Lambda-\frac{1}{2}(\rho_m+3 p_m+\rho_{{\rm phantom}}+3 p_{{\rm 
phantom}})
\ee
This is the second FRW equation for spatial curvature $k=0$ and using 
units for which $8\pi G=1$. 
\item[2.] {\it Physical Time Evolution generated by $H$: Modified FRW 
equations}\\
We can now compute the time evolution of physical observables. The 
general formula (\ref{3.1a}) specialises to
\be \label{4.25}
O_f(\tau)
=\sum_{n=0}^\infty \frac{1}{n!} \{H_\tau,f\}_{(n)}
=\sum_{n=0}^\infty \frac{(\tau-\phi)^n}{n!} \{H,f\}_{(n)},\;\;
H_\tau=(\tau-\phi)H
\ee
with 
\be \label{4.26}
H=\sqrt{(C+\beta a^3)^2-\alpha^2 a^6}
\ee
and $f$ can be any function on the cosmological minisuperspace phase space 
independent of $\pi,\phi$ because the spatial diffeomorphism constraint 
vanishes identically. 
Obviously, just as in the full theory $dO_f(\tau)=\{H,O_f(\tau)\}=
O_{\{H,f\}}(\tau)$. We now see that 
\be \label{4.27}
O_f(\tau)\approx f(\tau-\phi)\equiv f(t),\;\; t:=\tau-\phi
\ee
where $t\mapsto f(t),\;f(0)=f$ is the solution of the evolution equation 
for $f$
without the phantom when treating the Hamiltonian constraint $C$ as a 
Hamiltonian. The unphysical time parameter $t$ of that unphysical time 
evolution is now interpreted as $t=\tau-\phi$. Hence all the  
cosmological evolution equations remain approximately intact (under the 
restrictions on 
$c$ made above), however, we now have justified why that evolution 
corresponds to observation and we have interpreted the time parameter $t$
as composed of the pure gauge phantom time $\phi$ and the physical time 
parameter $\tau$.

Thus, the phantom has nicely reconciled the mathematical framework 
(gauge theory) with observation (FRW equations). It is pure gauge and one 
would be tempted to conclude that its presence does not have any 
observational consequences beyond modifying the Hamiltonian constraint. 
This is of course wrong because, at least 
in our model, we cannot have $C\equiv H$ just $C\approx H$ and we now 
proceed to compute the associated modifications. Hence, what we need to 
do is to repeat the steps (\ref{4.19}) -- (\ref{4.24}) where the first 
FRW equation $C^{{\rm tot}}=C+a^3 \rho_{{\rm 
phantom}}=0$ must be expressed in terms of observable quantities and 
instead of the gauge transformation $d/dt(.)=\{C^{{\rm 
tot}},.\}$ we now have actual physical evolution $d/d\tau(.)=\{H,.\}$.
The physical evolution equation for the observable $O_a$ 
corresponding to $a$ is\footnote{The convenient observation
here is that the map $f\mapsto O_f$ is an automorphism. In particular
if $f=F(a,P,\phi,\pi)$ for some function $F$ then 
$O_f=F(O_a,O_P,\tau,\pi)$.}  
\be \label{4.28}
\frac{d O_a}{d\tau}=\{H,O_a\}=O_{\{H,a\}}
=O_{\frac{\partial H}{\partial C} \{C,a\}}
=-O_{\frac{\partial H}{\partial C}} O_P/(6 O_a)
\ee
Here 
\be \label{4.29}
\frac{\partial H}{\partial C}=
\frac{C+\beta 
a^3}{H}=\epsilon\frac{\sqrt{E^2+\alpha^2 a^6}}{E}=\epsilon\sqrt{1+1/x}
\ee
Using (\ref{4.28}) in $O_{C^{{\rm tot}}}=0$ we find the first modified 
FRW 
equation\footnote{$O_{C^{{\rm tot}}}=C^{{\rm tot}}=0$ follows because 
$C^{{\rm tot}}$ is already gauge invariant. Here we need to use 
the generalisation of the map $f\mapsto O_f$ mentioned in section 
\ref{s3}.} 
\be \label{4.34}
\frac{3}{O_a^2}(\frac{dO_a}{d\tau})^2=(1+\frac{1}{O_x})[\Lambda+O_{\rho_m}+
O_{\rho_{{\rm phantom}}}]
\ee
The second again follows with the definition of pressure by taking the 
second physical time derivative of $O_a$
\be \label{4.35}
\frac{3}{O_a} \frac{d^2 O_a}{d\tau^2}=
(1+\frac{4}{O_x})\Lambda
-\frac{1}{2}\{[O_{\rho_m}+O_{\rho_{{\rm phantom}}}]
(1-\frac{5}{O_x})+3 [O_{p_m}+O_{p_{{\rm phantom}}}]
(1+\frac{1}{O_x}))\}
\ee
Notice that $O_x=E^2/(\alpha^2 O_a^6)$.
We now interpret (\ref{4.34}) and (\ref{4.35}): For $O_x\to \infty$ they 
look exactly like standard FRW equations with an additional phantom 
matter component. In fact, at $O_x=\infty$ these two equations are 
identical with (\ref{4.20}) and (\ref{4.24}) under the substitution
$(t,a(t))\to (\tau,O_a(\tau))$.
That additional matter component behaves at early times ($O_a\to 0$) of 
the universe ($O_x\to \infty$) as
\be \label{4.36}
\rho_{{\rm phantom}}\to -\epsilon E/a^3,\;\;p_{{\rm phantom}}\to -\beta
\ee
Hence it behaves like dust matter for $\epsilon=-1$ as $O_a\to 0$ 
provided we set $\beta=0$.
It therefore could serve as a dark matter candidate then. At later 
times,
when $O_a\to \infty$ or $O_x\to 0$ we get
\be \label{4.37}
\rho_{{\rm phantom}}\to \beta-\epsilon\alpha, 
\;p_{{\rm phantom}}\to -\beta+\alpha \epsilon
\ee
so it behaves like a positive/negative cosmological constant for 
$\epsilon=-1$ or $\epsilon=+1$ respectively. In order to get a 
positive contribution to the cosmological constant and to retain these
interpretations, we thus should choose
$\beta=0,\epsilon=-1$ which is exactly of the form used in the model 
\cite{K}. 

However, our purposes are different here: The phantom 
field is there in order to deparametrise the theory and to provide a 
positive, physical Hamiltonian which approximates $C$ since $C$ is used 
in the standard model with flat space. Since $C^{{\rm tot}}=C+\rho_{{\rm 
phantom}}=0$ we are forced to take $\rho_{{\rm phantom}}<0$ in order to 
have $C>0$. This excludes the $\epsilon=-1$ case\footnote{More precisely 
we could take $C^{\prime{\rm tot}}=\pi\pm \sqrt{C^2-\alpha^2 a^6}$, 
i.e. $H=\mp\sqrt{C^2-\alpha^2 a^6}\approx \pm |C|$ for small $a$ and 
$\alpha$. In order that this approximates $C$ we shoulld have $C<0$ 
or $C>0$ respectively which requires $\rho_{{\rm phantom}}>0$ or 
$\rho_{{\rm phantom}}<0$ respectively. However, 
we should have 
$\rho_{{\rm phantom}}<0$ in order to avoid a big rip singularity as 
we will show below. Also $C>0$ in flat space.} and we must use 
$\epsilon=+1$. Hence, the phantom has negative energy, positive pressure
and behaves like negative energy dust in the early universe while it 
apporaches a negative cosmological constant in the late universe. 
Since the gravitational contribution to $C$ is negative, we must have 
$\rho_m+\rho_{{\rm phantom}}>0$ in order that $C^{{\rm tot}}=0$. Hence 
the total energy density of matter
is positive which is sufficient in order to obtain a stable theory, 
that is, the usual energy conditions are satisfied \cite{Wald}. In 
fact the phantom can be compensated for by a k -- essence field with 
positive energy \cite{K} or by ordinary positive energy dust matter and 
a positive cosmological constant term. \\
\\
The reasoning here would then be as follows:\\
1. Something like a phantom is needed in order to deparametrise the 
theory and to keep validity of standard model physics and the FRW 
equations etc.\\
2. Since the phantom energy is negative we must compensate for it by 
positive energy matter. The simplest way to do this is to add a k -- 
essence field with energy density $\rho_k=\sqrt{\pi_k^2/a^6+\gamma^2}$
which just corresponds to an action of the DBI type with $\alpha$ 
replaced by $\gamma$ and $\epsilon=-1$. The k -- essence momentum 
$\pi_k$ is also a constant of the motion.\\
Thus, {\it both} the phantom and k -- essence fields are called for by 
the mathematical formalism.

Notice that the 
transition between the regimes where the usual FRW equations retain 
their interpretation as evolution equations of observable quantities is 
controlled by the deviation 
parameter $O_x=E^2/(\alpha^2  O_a^6)$. The 
transition occurs at at $O_x=1$ so $O_a=\root 3\of{E/\alpha}$. By 
choosing
$\alpha$ sufficiently small and/or $E$ sufficiently large we can 
achieve that the transition scale is 
as large as we need. Notice that $\alpha$ is a kinematical parameter 
of the Lagrangean while $E$ is a dynamical constant of motion. However, 
whatever the value of $\alpha$, equations 
(\ref{4.34}) differ drastically from the standard FRW equations beyond 
$O_x=1$. That is, the universe evolves completely differently beyond 
$O_x=1$. The largest corrections at late $O_x$ are of the order of 
$1/O_x \propto O_a^6$ which are terms normally not considered in the FRW 
equations. In fact they will completely dominate then.

Let us see what will qualitatively happen at very late times: We will
use realistic matter composed of dust (baryons), radiation and k -- 
essence (for simplicity without potential term) and set 
\be \label{4.37.0}
\rho_m=B/a^3+R/a^4+\sqrt{\pi_k^2/a^6+\gamma^2}
\ee
where $B,R>0$ are integration constants and $\gamma>0$. Notice that also 
$\pi_k$ is a constant of motion. The physical 
evolution 
equations become (we replace $O_a$ by $a$ etc. for the purpose of this 
discussion)
\be \label{4.37.1}
3(da/d\tau)^2/a^2=(1+1/x)[\Lambda+\rho_m+\rho_{{\rm phantom}}]
\ee
and we clearly need that the right hand side is positive for the entire
evolution. During radiation domination, $R>0$ is sufficient. During 
baryon domination we must have $B+|\pi_k|-|E|>0$. For $x\to \infty$ 
we may want to require $\Lambda+\gamma-\alpha\ge 0$. However, we notice 
one problem at late times:
Suppose that the right hand side of (\ref{4.37.1}) never vanishes. 
Then $da/d\tau>0$ for all $\tau$ due to continuity. Expanding 
(\ref{4.37.1}) around $x=0$ we find 
\be \label{4.37.2}
3(da/d\tau)^2/a^2=(1+1/x)\{[\Lambda+\gamma-\alpha]+
\frac{1}{2}(\gamma \frac{\pi_k^2 \alpha^2}{E^2 
\gamma^2} -\alpha) x+B \frac{\alpha}{E} \sqrt{x}+ R(\frac{\alpha}{E} 
\sqrt{x})^{4/3} +O(x^2)\}
\ee
Even if $\Lambda+\gamma-\alpha=0$ the right hand side diverges at $x\to 
0$ due to the radiation and baryonic terms present. The leading 
contribution is then given by 
\be \label{4.37.3}
(da/d\tau)^2=B\frac{\alpha^2}{3E^2} a^5=:\kappa^2 a^5
\ee
which can be solved by 
\be \label{4.37.4}
a(\tau)=\frac{1}{(\delta\mp \frac{3}{2} \kappa \tau)^{2/3}}
\ee
with $\kappa>0$, $\delta$ is a constant of integration and the 
upper/lower sign corresponds to $da/d\tau>0/<0$ respectively. Hence,
if $da/d\tau>0$ for all $\tau$ then we must take the upper sign and 
$\delta>0$ and would conclude that the universe reaches {\it infinite} 
size after the {\it finite} amount of time $2 \delta/(3\kappa)$. The 
evolution would in fact stop there and is called a big rip singularity. 
This 
is clearly undesirable and the only way to avoid this is to tune the 
parameters in such a way that $\rho$ can vanish while always being non - 
negative during the evolution. This is only possible if the phantom 
contribution to the energy budget is negative. 
The universe would then be able to reach maximum size and then
would recollapse. This is granted to happen if the right hand side 
of (\ref{4.37.1}) kinematically can become negative beyond some 
critical $x_c$ (dynamically it can never happen because $(da/d\tau)^2\ge 
0$). In fact it is not difficult to show that we must have 
$\Lambda+\gamma-\alpha<0$ for this to happen. 

One  
can tune $B,R,\Lambda,E,\pi_k,\gamma,\alpha$ such that there is a
radiation, dust and positive vacuum energy era while still $x\gg 1$ 
during which the FRW equations hold. The radiation era holds for 
$0\le a\le \frac{R}{B+|\pi_k|-E}=:a_r$ during which 
$\sqrt{x}\ge \frac{E}{\alpha a_r^3}\gg 1$ must 
hold. The dust era is $a_r\le a \le \root 3 
\of{\frac{B+|\pi_k|-E}{\Lambda}}=:a_d$ during which 
$\sqrt{x}\ge \frac{E}{\alpha a_d^3}\gg 1$ must hold. 
Since $\Lambda/\alpha<1$ and $a_r<a_d$, both conditions are easily 
satisfied with the observed values for $R,B$ for 
$|\pi_k|-E$ of the order of 
$B$, $E/B\gg 1$ and $E/\alpha$ sufficiently small. Finally, the 
vacumm 
era lasts for $a_d\le a \le a_c$ where $a_c$ is the value at which 
(\ref{4.37.1}) vanishes. In order to see that $x_c$ is smaller than 
$x=1$ at which the FRW equations anyway no longer take their standard 
form we notice that at $x=1$ the right hand side of (\ref{4.37.2}) is
still larger than 
$\Lambda+\sqrt{\gamma^2+\alpha^2\frac{\pi_k^2}{E^2}}-\sqrt{2}\alpha$
which is positive independently of the value of $\Lambda,\gamma$ as long 
as $\pi_k^2/E^2\ge 2$. However, if we associate $B$ with baryonic matter 
and $|\pi_k|-E$ with dark matter then $(|\pi_k|-E)/B\approx 10$. Since 
$B/E \ll 1$ we get $|\pi_k|/E\approx 1$ so we should have 
$\Lambda+\sqrt{\gamma^2+\alpha^2}-\sqrt{2}\alpha>0$ for 
$\Lambda,\gamma,\alpha >0$ subject to $\Lambda+\gamma<\alpha$.
Fix some $\epsilon>0$ and set $y:=\gamma/\alpha$ and 
$\Lambda/\alpha:=(1-\epsilon-y)$. Then we need to find $0\le y\le 
1-\epsilon$ 
such that $1-\epsilon-y+\sqrt{1+y^2}-\sqrt{2}\ge 0$. Set 
$\kappa:=\sqrt{2}-1+\epsilon$ which is positive and satisfies
$\kappa<1$ for $\epsilon<2-\sqrt{2}$. Then 
$0\le y=(1-\kappa^2)/(2\kappa)\le 1-\epsilon$ which implies 
$1+(1-\epsilon)^2\le 2$ which is identically satisfied for $0<\epsilon\le 
2$. Hence we may choose any $0<\epsilon\le 2-\sqrt{2}$ and then 
$\gamma=\alpha (1-\kappa^2)/2\kappa,\;\Lambda=\alpha(1-\epsilon)-\gamma$ 
where $\kappa=\sqrt{2}-1+\epsilon$. It is appealing that the 
cosmological constant during the vacuum era $\Lambda$ is of the order of 
$\alpha$ which should be small, thus explaining the smallness of the 
cosmological constant\footnote{For completeness we also mention a 
scenario without k -- essence matter, that is, $\pi_k=\gamma=0$. In this 
case the same analysis yields the condition $0<E/(B-E)\gg 1$. Then 
at $x=1$ the energy density becomes $\rho\ge 
\Lambda+\alpha(B/E-\sqrt{2})$ which is still positive for 
$\sqrt{2}-1-\epsilon\le \Lambda/\alpha <1$ where $B=(1+\epsilon)E$. Notice 
that now $B-E$ corresponds to observed baryonic matter.}
\end{itemize}
Let us summarise once again the observable effect of the phantom:\\
1. The physical evolution equations of observable quantities have 
a standard FRW form for large $x$ (small $a$).\\
2. The phantom adds additional matter terms with an equation of state 
$-1\le w\le 0$ which evolves from $0 \to -1$ as $a$ evolves from $0\to 
\infty$. It therefore acts like in k -- essence, just with negative 
energy. However, an additional k -- essence field is natural
in order to compensate the negative phantom energy.\\
3. In particular, it is wrong that the physical evolution equations
have just the FRW form without phantom matter contribution. It is easy 
to see that this is due to the fact that $C^{{\rm 
tot}}\not=C^{{\rm ideal}}:=C+\pi$ in 
which case we would have $H=C$ exactly.
Rather we have 
$C^{{\rm tot}}=C-\sqrt{\pi^2+\alpha^2 a^6}$ so that $C^{{\rm 
tot}}-C^{{\rm 
ideal}}=|\pi|(\sqrt{1+1/x}-1)$. It 
is not difficult to see that there is 
no Lagrangean of the form $L(I/2)$ which can produce $H=C$ exactly.\\
4. As $x$ becomes small, the actual evolution equations differ 
drastically from the FRW form. The transition is roughly at 
$a_t=\root 3 \of{E/\alpha}$ where $E$ is the nergy of the universe and 
$\alpha$ is a parameter of the model which can be tuned to be so small 
that $a_t$ 
is way beyond today's value $a_0$. In order that the universe has 
infinite observable life time, the parameters can and must be tuned such 
that the universe in fact recollapses rather than expanding forever.\\
\\
Notice that we are not doubting the validity of Einstein's equations at 
all. These are completely encoded in the fundamental constraint $C^{{\rm 
tot}}$. 
There are two of these equations. One is $C^{{\rm tot}}=0$. The 
other results by computing the gauge transformation 
$da/dt=\{C^{{\rm tot}},a\}$, to solve this for $P$, to compute the 
gauge transformation $dP/dt=\{C^{{\rm tot}},P\}$ and to insert this
into $d^2a/dt^2$. This results exactly in (\ref{4.34}), (\ref{4.35}) at 
$x=\infty$
and with $t$ replaced by $\tau$. However, what we critisise is that 
these are interpreted as physical evolution equations. They are 
not, they are gauge transformations of non -- gauge invariant, 
unobservable
quantities and not evolution equations with respect to a non -- 
vanishing Hamiltonian of observable quantities. What we have done here 
is to compute the physical evolution of observable quantities 
generated by a physical Hamiltonian. The resulting equations are,
by a judicious choice of phantom field Lagrangean, in good agreement 
with the standard equations. However we insist that the standard 
procedure is fundamentally wrong. In 
particular, the standard FRW equations are drastically {\it false} in 
the late universe if our phantom field is realised in nature and 
provides the physical Hamiltonian which generates the evolution that we 
actually observe.
These reservations hold of course in full generality in all applications 
of General Relativity. \\
\\
There is another way to look at what is going on here:
What we have done in order to obtain physical evolution is to use the 
unphysical gauge transformation 
$d\phi/dt(t)=\{C^{{\rm tot}},\phi\}_{\phi=\phi(t)}$ and to solve the 
condition
$\phi(t)=\tau$ for $t$. This results in the function $\tau\mapsto 
t_\tau$ which is a non  -- trivial function on phase space. Now we 
insert the value $t_\tau$ into the unphysical gauge transformation
$t\mapsto a(t)$ where $a(t)$ solves 
$da/dt(t)=\{C^{{\rm tot}},t\}_{a=a(t)}$ and obtain $a(t_\tau)$. We claim
that $a(t_\tau)=O_a(\tau)$, at least when $C^{{\rm tot}}=0$, where
$O_a(\tau)$ solves 
$dO_a/d\tau(\tau)=\{H,O_a(\tau)\}$. To see this, we compute 
for any 
function $f$ with equation of motion $df/dt=\{C^{{\rm tot}},f\}$ 
\be \label{4.38}
\{C^{{\rm tot}},f(t_\tau)\}=
\{C^{{\rm tot}},f\}_{f=f(t),t=t_\tau}+(df/dt)_{t=t_\tau} 
\{C^{{\rm tot}},t_\tau\}
=(df/dt)_{t=t_\tau}[1+\{C^{{\rm tot}},t_\tau\}]
\ee
Now choose $f=\phi$ in (\ref{4.38}) and use that $\tau=\phi(t_\tau)$ is 
a constant function. Then use $f=a$ in (\ref{4.38}) to see that 
$a(t_\tau)$ is an observable. Now on the constraint surface  
\be \label{4.39}
0=\{C^{\prime{\rm tot}},f(t_\tau)\}=
\{\pi,f\}_{f(t),t=t_\tau}+\{H,f\}_{f(t),t=t_\tau}+
(df/dt)_{t=t_\tau}[\{\pi,t_\tau\}+\{H,t_\tau\}]
\ee
Choose $f=\phi$ and use that $H$ does not depend on $\pi$, then
\be \label{4.40}
0=1+(d\phi/dt)_{t=t_\tau}[\{\pi,t_\tau\}+\{H,t_\tau\}]
\ee
Insert (\ref{4.40}) into (\ref{4.39}) with $f=a$ to obtain
\be \label{4.41}
0=\{H,a\}_{a=a(t),t=t_\tau}-
(da/dt)_{t=t_\tau}/[(d\phi/dt)_{t=t_\tau}]
=\{H,a\}_{a=a(t_\tau)}-da(t_\tau)/d\tau
\ee
Thus $a(t_\tau)$ and $O_a(\tau)$ differ at most by a constant.

What happens here is that $t\mapsto a(t),\;t\mapsto \phi(t)$ 
is the parametrisation of a trajectory in phase 
space (here restricted to the $a,\phi$ plane) which is obtained by 
solving the equation $\phi(t)=\tau$ for $t$ and inserting this into 
$a(t)$ so that we arrive at $\tau\mapsto a(t_\tau)$. We have {\it 
deparametrised} the description and now are dealing with the only 
physically meaningful object, the trajectory itself and not some random 
parametrisation thereof. Any reparametrisation $t=t(t')$ with 
$dt/dt'>0$, that is, a gauge transformation, changes the unphysical 
functions $r(t),\phi(t)$ but results in the same physical trajectory.  

What consequences does this have for the FRW line element
\be \label{4.44}
ds^2=-dt^2+a(t)^2 dx^a dx^b \delta_{ab}
\ee
which is also expressed in terms of the unphysical quantities $t,a(t)$?
Let us express the line element in terms of $\tau$ by applying the 
diffeomorphism $t:=t_\tau$. In these coordinates (\ref{4.44}) becomes
\be \label{4.45}
ds^2=-d\tau^2 [\frac{d\phi}{dt}(t)]^2_{t=t_\tau} +a(t_\tau)^2 dx^a dx^b 
\delta_{ab}
=-d\tau^2 (1+\frac{\alpha^2 a(t_\tau)^6}{E^2}) +a(t_\tau)^2 dx^a dx^b 
\delta_{ab}
\ee
which is again a FRW line element, now expressed in terms of observable 
quantities, at least for small $a$. For large $a$, (\ref{4.45}) is no 
longer of FRW form. Again the deviation parameter $x$ has appeared and 
shows that for sufficiently small $a$ the metric (\ref{4.45}) expressed in
terms of observable quantities is well approximated by the usual FRW 
form even today.

\section{Conclusions and Outlook}
\label{s5}

It has been known for a long time that the problem of time can be solved 
in principle by the relational framework due to Rovelli and others.
This has never been appreciated as much as it should have been because,
while the conceptual, physical framework was clear, the analytical 
implementation remained largely undeveloped for a long time. With 
the appearance of \cite{Bianca}, analytical methods became available for 
the first time. Still the framework, in its full generality, remains 
discouragingly difficult in particular when applied
to General Relativity due to the complexity of the analytical expressions 
which involve summing an infinite number of infinite series, an inversion 
of infinite dimensional matrices and the computation of an infinite number 
of different, iterated Poisson brackets\footnote{Drastic simplifications 
occur as far as the number of relevant constraints is concerned (four 
instead of infinitely many) when 
reformulating GR in terms of coordinates that are spacetime scalars 
\cite{Bianca1}. While the resulting observables are relatively simple
(although still inversions of non -- trivial matrices take place) and 
physically intuitive, the field variables that one uses are complicated 
compounds of the the canonical fields and the observables involve 
polynomials of those evaluated at one specific spatial point. In quantum 
theory, these observables are therefore presumably too singular 
because they involve the product of operator valued distributions 
evaluated at the same point.}.

The first main message of the present paper is that when adding 
appropiate, albeit hypothetical, matter, the complexity of these formulas 
is drastically reduced. In contrast to the general case, there is only one 
series to sum, there are no matrices to invert, there is only one kind of 
iterated Poisson bracket to compute. Hence the formulas that we obtain are 
remarkably simple. In fact, the classical time evolution in a background 
dependent theory, 
say in QCD on Minkowski space, of 
some observable $O$ such as a Wilson loop function,
would also be given by the series
\be \label{5.0}
O(\tau)=\sum_{n=0}^\infty \frac{\tau^n}{n!} \{H_{{\rm QCD}},O\}_{(n)}
\ee
where $H_{{\rm QCD}}$ is the QCD Hamiltonian. Comparing with (\ref{3.1a})
we see that the complexity of the classical time evolution in both 
theories is comparable! 

The second main message is that,  
in contrast to the general case, the physical 
observables we obtain are strong observables and there is just one 
natural, physical Hamiltonian which does not depend on the 
physical time parameter. That Hamiltonian is (constrained to be) positive 
and at least in the physically interesting 
region in phase space, that Hamiltonian reduces to the canonical 
Hamiltonian that one usually uses in cosmology and the standard model
when the metric is flat.
In fact, we manage to completely 
deparametrise the system irrespective of the other matter present.

The third main message is that the scalar type of matter that we 
considered here, from the mathematical (to be able to solve 
algebraic equations) and physical (to obtain a physical Hamiltonian 
which is close to that of the standard model) perspective naturally leads 
to Dirac -- Born -- Infeld (DBI) negative energy phantom fields with 
constant potential. This negative energy phantom must be compensated for 
by positive energy matter, most naturally by a k -- essence field.
Such matter was discussed independently in the cosmology literature in 
order to provide 
a candidate for inflation and dark energy. Hence the scalar matter we 
consider here might actually really exist! 

The fourth main message is that, at least for the scalar model we have 
used here and for which we gave strong motivations, the usual 
interpretation of the cosmological framework, although fundamentally 
wrong because gauge transformations of gauge dependent objects are 
interpreted as actual physical evolution equations of observables,
remains valid when analysed in the correct way, that is, by computing 
the physical evolution of gauge invariant observables. The 
domain of validity of these equations can be tuned to be arbitrarily 
long, however, it is manifestly finite when using the physical time 
parameter corresponding to the physical Hamiltonian. The actual 
evolution at late times apparently leads to a recollapse.\\
\\
The future application of this framework lies of course in the quantum 
theory. The framework presented here, as well as any other 
application of the relational Ansatz so far, is purely classical. In 
order to promote the framework to the quantum theory, 
the functions $H$ should 
be promoted to positive self -- adoint operators and the 
functions $H_\tau$ and $H(M)$ to self --adjoint 
operators\footnote{These operators are supposed to be spatially 
diffeomorphism invariant, see \cite{LQG} for a quantum implementation of 
the diffeomorphism group within LQG.}.
If we find operator orderings such that $[\hat{H}(M),\hat{H}(M')]=0,\;
[\hat{H}(M),\hat{H}_\tau]=i\hbar\hat{H}(M)$ then the quantum observables 
are given by
\be \label{5.1}
\widehat{O_f}(\tau)=\exp(i\hat{H}_\tau/\hbar) \hat{f} 
\exp(-i\hat{H}_\tau/\hbar)
\ee
where the unitary operators displayed are defined by the spectral theorem.
They manifestly commute, under the assumptions made, with 
$\exp(i\hat{H}(M)/\hbar)$ and the operator ordering problem for the 
observables would be solved\footnote{If one cannot find a model in which 
all expressions of which one has to take the square root are manifestly 
positive, then we may be able to compute the spectrum of the 
(regulated) operators 
without the square root and restrict the Hilbert space to the 
``subspace'' on which all of them take positive (generalised) 
eigenvalues. This is possible because the operators are supposed 
to commute. The square root would then be well defined on that subspace as 
has 
been pointed out in \cite{BK}. Alternatively one can try to use 
the manifestly positive substitute expressions (\ref{4.8c}).}

Given these assumptions, the fact that then a positive, fundamental 
Hamiltonian is available could enable one to solve the vacuum problem 
in quantum cosmology: Namely, in usual semiclassical quantum cosmology one 
neglects quantum gravity and applies the framework of quantum field theory 
on curved spacetimes \cite{Wald1}. The issue is that in cosmology the 
background metric is not stationary and therefore the problem becomes,
roughly speaking, to choose a point of unphysical time and at that time a
definition of annihilation operators (for the free fields) suggested by 
the energy density function of the matter in question in order to 
select a vacuum state. This is highly ambiguous and the cosmological 
evolution does not keep the vacuum intact but rather causes constant 
particle production. On the other hand, if one has a fundamental 
Hamiltonian at one's disposal, then it is natural to define a vacuum state 
as a minimal (zero) energy (eigen)state. This would circumvent this 
problem of initial conditions. 

Having a physical Hamiltonian and physical observables at one's disposal 
one can also hope to develop physical scattering and S -- Matrix theory. 
Namely, while we drastically simplified the relational framework, it will 
be still
very hard to compute $\widehat{O_f}(\tau)$ explicitly to all orders. 
Here one will use the series in order to perform perturbation theory
in the way outlined in \cite{BT}, say within the framework of LQG:
Given approximate physical states which can be obtained by using 
semiclassical techniques of LQG \cite{GCS}, we can concentrate them 
on regions in phase space where $\phi(x)\approx \tau$. Then expectation 
value computations of physical observables can be terminated after a 
few terms in the power series and only a small number of iterated 
commutators has to be computed. This should work especially nice in 
applications to quantum cosmology \cite{BT} within LQG.\\
\\
In summary, there is much left to do in order to make this framework 
practically applicable and it is worthwhile to explore the space of 
Lagrangeans which lead to deparametrisation further. However, we feel 
that conceptually the framework 
is quite clear, the complexity has been drastically reduced, its 
validity has been checked in a cosmological setting and the 
remaining technical tasks to be solved have been 
identified.\\
\\
\\
\\
{\large Acknowledgements}\\
\\
We thank Kristina Giesel, Stefan Hofmann and Oliver Winkler for inspiring 
discussions and for making valuable suggestions for improvement of the 
original manuscript. Special thanks go to Stefan Hofmann for a critical 
reading of the article from the point of view of a cosmologist and for 
pointing out the relevant literature.
This research project was supported in part by a 
grant from NSERC of Canada to the Perimeter Institute for Theoretical Physics.


\end{document}